\documentclass{aastex631}
\usepackage{CJK}
\usepackage{units}
\usepackage{gensymb}
\usepackage{subfigure}
\usepackage[percent]{overpic}

\graphicspath{{./}{figures/}}

\begin{document}
\begin{CJK*}{UTF8}{gbsn}

\title{The role of binarity and stellar rotation in the split main sequence of NGC 2422}

\correspondingauthor{Chengyuan Li, Weijia Sun}
\email{lichengy5@mail.sysu.edu.cn,weijia@gmail.com}

\author[0000-0001-9131-6956]{Chenyu He (贺辰昱)}
\affiliation{School of Physics and Astronomy, Sun Yat-sen University, Zhuhai, 519082, China}
\affiliation{CSST Science Center for the Guangdong--Hong Kong--Macau Greater Bay Area, Zhuhai, 519082, China}

\author[0000-0002-3279-0233]{Weijia Sun (孙唯佳)} 
\affiliation{Department of Astronomy, Peking University, Yi He Yuan Lu 5, Hai Dian District, Beijing, 100871, China}
\affiliation{CAS Key Laboratory of Optical Astronomy, National Astronomical Observatories, Chinese Academy of Sciences, Beijing 100101, China}

\author[0000-0002-3084-5157]{Chengyuan Li (李程远)}
\affiliation{School of Physics and Astronomy, Sun Yat-sen University, Zhuhai, 519082, China}
\affiliation{CSST Science Center for the Guangdong--Hong Kong--Macau Greater Bay Area, Zhuhai, 519082, China}

\author[0000-0002-0880-3380]{Lu Li (李璐)}
\affiliation{Key Laboratory for Research in Galaxies and Cosmology, Shanghai Astronomical Observatory, Chinese Academy of Sciences, 80 Nandan Road, Shanghai, 200030, China}
\affiliation{School of Astronomy and Space Science, University of Chinese Academy of Sciences, 19A Yuquan Road, Beijing, 100049, China}
\affiliation{Centre for Astrophysics and Planetary Science, Racah Institute of Physics, The Hebrew University, Jerusalem, 91904, Israel}

\author[0000-0001-8611-2465]{Zhengyi Shao (邵正义)}
\affiliation{Key Laboratory for Research in Galaxies and Cosmology, Shanghai Astronomical Observatory, Chinese Academy of Sciences, 80 Nandan Road, Shanghai, 200030, China}
\affiliation{Key Laboratory for Astrophysics, Chinese Academy of Sciences, Shanghai 200234, China}

\author[0000-0001-5245-0335]{Jing Zhong (钟靖)}
\affiliation{Key Laboratory for Research in Galaxies and Cosmology, Shanghai Astronomical Observatory, Chinese Academy of Sciences, 80 Nandan Road, Shanghai, 200030, China}

\author[0000-0002-4907-9720]{Li Chen (陈力)}
\affiliation{Key Laboratory for Research in Galaxies and Cosmology, Shanghai Astronomical Observatory, Chinese Academy of Sciences, 80 Nandan Road, Shanghai, 200030, China}
\affiliation{School of Astronomy and Space Science, University of Chinese Academy of Sciences, 19A Yuquan Road, Beijing, 100049, China}

\author[0000-0002-7203-5996]{Richard de Grijs}
\affiliation{School of Mathematical and Physical Sciences, Macquarie University, Balaclava Road, Sydney, NSW 2109, Australia}
\affiliation{Research Centre for Astronomy, Astrophysics and Astrophotonics, Macquarie University, Balaclava Road, Sydney, NSW 2109, Australia}

\author[0000-0002-0066-0346]{Baitian Tang (汤柏添)}
\affiliation{School of Physics and Astronomy, Sun Yat-sen University, Zhuhai, 519082, China}
\affiliation{CSST Science Center for the Guangdong--Hong Kong--Macau Greater Bay Area, Zhuhai, 519082, China}

\author[0000-0003-3713-2640]{Songmei Qin (秦松梅)}
\affiliation{Key Laboratory for Research in Galaxies and Cosmology, Shanghai Astronomical Observatory, Chinese Academy of Sciences, 80 Nandan Road, Shanghai, 200030, China}
\affiliation{School of Astronomy and Space Science, University of Chinese Academy of Sciences, 19A Yuquan Road, Beijing, 100049, China}

\author[0000-0003-2666-4158]{Zara Randriamanakoto}
\affiliation{South African Astronomical Observatory, P.O. Box 9, Observatory, Cape Town 7935, South Africa}
\affiliation{Department of Physics, University of Antananarivo, P.O. Box 906, Antananarivo, Madagascar}

\date{August 2022}

\begin{abstract}
In addition to the extended main-sequence turnoffs widely found in
young and intermediate-age ($\sim\unit[600]{Myr}$--$\unit[2]{Gyr}$-old) star clusters,
some younger clusters even exhibit split main sequences
(MSs). Different stellar rotation rates are proposed to account for
the bifurcated MS pattern, with red and blue MSs (rMS and bMS)
populated by fast and slowly rotating stars, respectively. Using
photometry from \textit{Gaia} Early Data Release 3, we report a
Galactic open cluster with a bifurcated MS, NGC 2422 ($\sim
\unit[90]{Myr}$). We exclude the possibilities that the bifurcated MS
pattern is caused by photometric noise or differential reddening. We
aim to examine if stellar rotation can account for the split MS. We
use spectra observed with the Canada--France--Hawaii Telescope and
the Southern African Large Telescope, and directly measured $v\sin i$,
the projected rotational velocities, for stars populating the bMS and
rMS. We find that their $v\sin i$ values are weakly correlated with
their loci in the color--magnitude diagram because of contamination
caused by a large fraction of rMS stars with low projected rotational
velocities. Based on the spectral energy distribution fitting method,
we suggest that these slowly rotating stars at the rMS may hide a
binary companion, which breaks the expected $v\sin i$--color
correlation. Future time-domain studies focusing on whether these
slowly rotating stars are radial velocity variables are crucial to test
the roles of stellar rotation and binarity in generating the split
MSs.
\end{abstract}


\section{introduction\label{sec:intro}}

Studies based on Hubble Space Telescope (\textit{HST}) and
\textit{Gaia} observations have revealed that many star clusters
younger than 2 Gyr exhibit extended main-sequence turnoffs
\citep[eMSTOs; e.g.,][]{2008ApJ...681L..17M, 2009A&A...497..755M,
  2018MNRAS.477.2640M,2018ApJ...869..139C,
  2019ApJ...876...65L}. Clusters younger than $\unit[600]{Myr}$ also present split main
sequences \citep[MSs;][]{2016MNRAS.458.4368M, 2018MNRAS.477.2640M,
  2017MNRAS.467.3628C, 2017ApJ...844..119L, 2019ApJ...883..182S}. The
presence of eMSTOs and split MSs in clusters has challenged the
scenario that clusters are `simple stellar populations' whose
color--magnitude diagrams (CMDs) can be described by a single
isochrone characterized by a unique age and metallicity. The eMSTOs
were initially thought to be the result of an extended star formation
history (eSFH) within the clusters
\citep[e.g.,][]{2009A&A...497..755M,2009AJ....137.4988G,2011ApJ...737....4G},
lasting 150--500 Myr in intermediate-age clusters. This scenario,
however, attracted lots of objections, such as regarding the absence
of gas in Magellanic Cloud (MC) young massive clusters
\citep[YMCs;][]{2014MNRAS.443.3594B}, and relating to the morphologies
of the sub-giant branches (SGBs) and/or red clumps (RCs) in some
massive clusters disfavoring an eSFH
\citep[][]{2014ApJ...784..157L,2014Natur.516..367L,2016MNRAS.461.3212L,
  2015MNRAS.448.1863B}. Another scenario attributes eMSTOs and split
MSs to different stellar rotation rates
\citep[e.g.,][]{2009MNRAS.398L..11B,
  2018MNRAS.477.2640M,2018ApJ...869..139C,2019ApJ...876..113S,
  2019ApJ...883..182S}. The reduction in self-gravity due to rotation
will reduce the stellar surface temperature, leading stars to appear
redder than their slowly rotating counterparts
\citep{1924MNRAS..84..665V}. This so-called `gravity darkening' effect
works more significantly at stellar equators than at their poles, and
thus the observed color of a fast rotating star also depends on its
inclination. Another effect, rotational mixing, could expand the
stellar convective core, prolonging the MS lifetime of fast rotating
stars \citep[][]{2000ARA&A..38..143M}. Both effects, caused by stellar
rotation, will complicate the morphologies of the MS and MSTO regions
of a `simple stellar population' cluster.

Only MS dwarfs less massive than $\sim 1.5 M_\odot$ are slow rotators
because of the magnetic braking effect
\citep[][]{1967ApJ...150..551K}. Because massive MS dwarfs cannot
survive in older clusters, if eMSTOs and split MSs are caused by
stellar rotation, then only clusters younger than a critical age would
exhibit eMSTOs and split MSs. This was confirmed by
\citet{2018MNRAS.477.2640M}, who found that only MS stars more massive
than $1.6 M_\odot$ could populate a bifurcated MS. For the eMSTO
phenomenon, spectroscopic studies of the Large Magellanic Cloud (LMC) clusters NGC 1866 ($\sim
\unit[200]{Myr}$) and NGC 1846
($\sim \unit[1.5]{Gyr}$), and the Galactic open cluster NGC 5822 ($\sim \unit[0.9]{Gyr}$), have revealed that the average projected
rotational velocity, $v\sin i$, of stars on the red side of the MS is
higher than that of stars on the blue side \citep{2017ApJ...846L...1D,2020MNRAS.492.2177K,2019ApJ...876..113S}. In
addition, a significant fraction of Be stars are detected around the
MSTO regions of some young MC clusters
\citep[e.g.,][]{2018MNRAS.477.2640M,2017MNRAS.465.4795B}, and these Be
stars are much redder than normal stars near the MSTO region
\citep{2018MNRAS.477.2640M}.

The split MSs can be ideally reproduced by two coeval stellar
populations with different rotational rates, i.e., fast/slow rotating
stars populate the red/blue sequences \citep[][]{2015MNRAS.453.2637D,
  2016MNRAS.458.4368M, 2017NatAs...1E.186D}. Additional evidence comes
from spectroscopic analyses of clusters in the LMC and the Milky Way. \citet{2018AJ....156..116M} found that
the blue and red MS (bMS and rMS) stars in the 40 Myr-old YMC NGC 1818 in the LMC exhibit an obvious difference in $v\sin i$. This bimodality of $v\sin i$ is detected among the eMSTO stars of the intermediate-age LMC cluster
NGC 1846 ($\sim\unit[1.5]{Gyr}$) \citep{2020MNRAS.492.2177K} as
well. Using photometry and high-resolution spectra, \citet{2018ApJ...863L..33M} found that a nearby young ($\sim \unit[300]{Myr}$-old) cluster NGC 6705 exhibits an eMSTO and a broadened MS, and its bMS and rMS are populated by slowly and fast rotating stars. \citet{2019ApJ...883..182S} analyzed the $v\sin i$ distribution along the well-separated MS of a young Galactic cluster NGC 2287 ($\sim\unit[200]{Myr}$), and found it shows strong evidence of a bimodal distribution of stellar absolute rotation rates.

The origin of the dichotomy in stellar rotation rates is yet to be
examined. To fully explain the morphology of eMSTOs, an age difference
between two populations with different rotation rates is required
\citep[][]{2016MNRAS.458.4368M,2017MNRAS.465.4363M,2017MNRAS.467.3628C,2017ApJ...846...22G,2018ApJ...864L...3G,2019ApJ...876...65L}. \citet{2017NatAs...1E.186D} suggested that
stars on the bMS could be produced by slow rotators that were
initially rapidly rotating stars and then were braked within a
timescale of less than 25\% of the cluster age. This also helps to
explain the apparent age differences among eMSTO stars. They proposed
that tidal locking in close binaries is responsible for the
braking. On the other hand, \citet{2020MNRAS.495.1978B} argued that
the bimodal distribution of stellar rotational velocities might
manifest itself at the pre-main-sequence (PMS) stage, possibly
introduced by different disk-locking mechanisms during the PMS
phase. Recently, \citet{2021MNRAS.508.2302K} detected similar binary
fractions in the slow and rapidly rotating stars along the split MS of
the massive LMC cluster NGC 1850 ($\sim \unit[100]{Myr}$). Combined
with the similar binary fractions detected in the slow and fast
rotating stars at the eMSTO of NGC 1846 by
\citet{2020MNRAS.492.2177K}, \citet{2021MNRAS.508.2302K} argue that
bimodal rotational distributions are not predominantly caused by
binary-induced braking. Recently, \citet{2022NatAs.tmp...35W}
suggested a binary-merger origin for the formation of slow rotators on
the blue MS, which is reflected by the unusual mass functions of the
bMS stars they detected in NGC 1755, NGC 330, NGC 1818, and NGC 2164.

In this paper, we report on another Galactic open cluster, NGC 2422,
which exhibits a bifurcated MS. It cannot be explained by photometric
uncertainties or differential reddening, which leaves stellar rotation
or/and unresolved binaries as possible explanations. The bifurcated MS
of NGC 2422 thus provides a rare opportunity to constrain the effects
of stellar rotation and binarity. We studied the eMSTO and split-MS
stars of NGC 2422 based on \textit{Gaia} Early Data Release 3
\citep[EDR3;][]{2016A&A...595A...1G,2021A&A...649A...1G}. We obtained
and resolved high- and medium-resolution spectra of individual stars
along the split MS. The aim of this study is to examine whether stars
along the cluster's split MSs also follow a similar correlation
between their $v\sin i$ and photometric colors as seen for NGC 2287
\citep{2019ApJ...883..182S}.

This article is organized as follows. In Section~\ref{sec:data} we
describe the data reduction. The main results, as well as their
scientific implications are presented in Section~\ref{sec:discu}. In
Section~\ref{sec:conc}, we briefly discuss our results and summarize
our conclusions.

\section{Data Reduction\label{sec:data}}
\subsection{Member Star Selection}

\begin{figure}
 \centering
 \subfigure{\label{fig:pm}
  \begin{overpic}[scale=0.47]{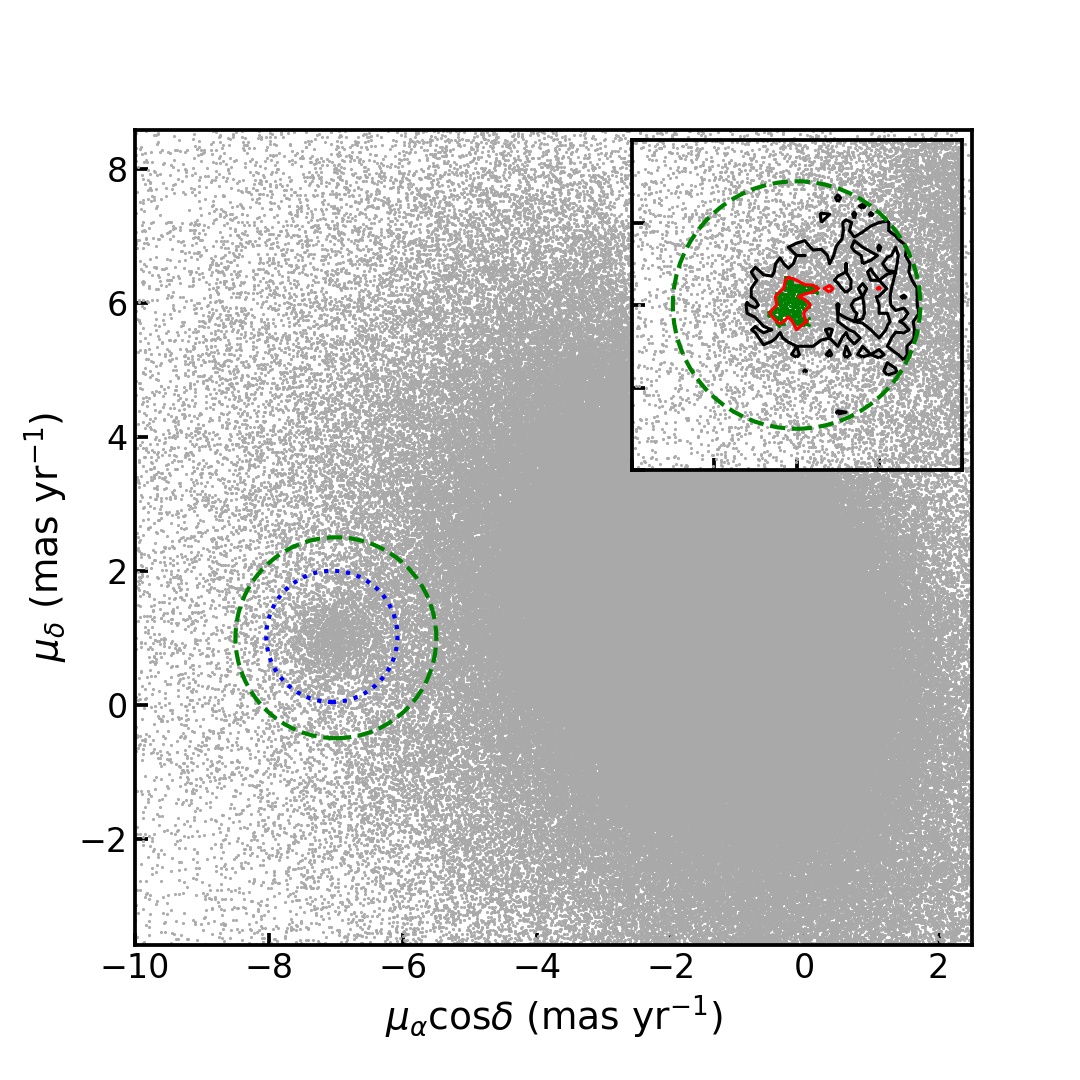}
    \put(20,80){(a)}
  \end{overpic}}
  \hspace{0.005\linewidth}
 \subfigure{\label{fig:para}
  \begin{overpic}[scale=0.47]{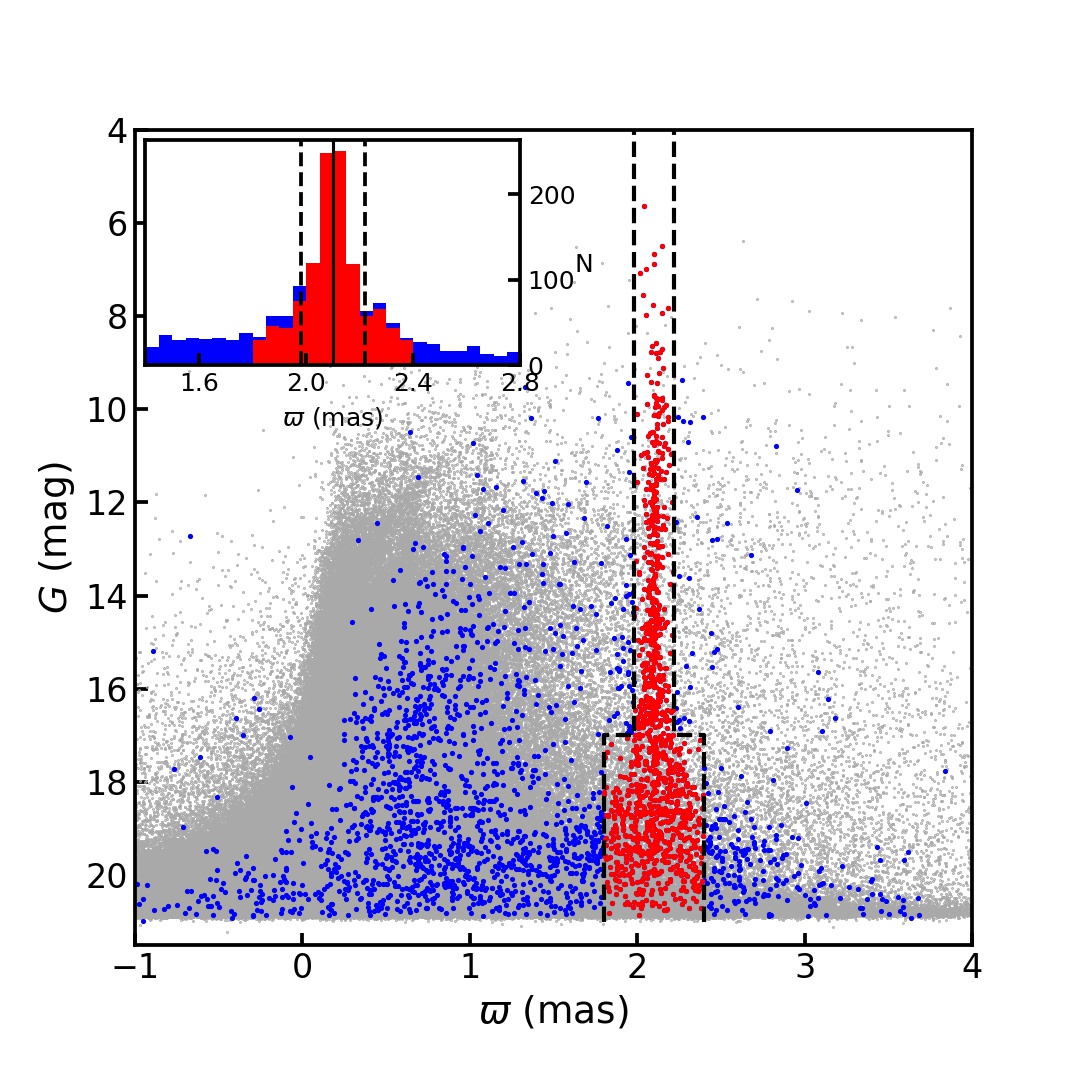}
   \put(80,80){(b)}
  \end{overpic}}
  \vfill
 \subfigure{\label{fig:cmd_allstar}
  \begin{overpic}[scale=0.47]{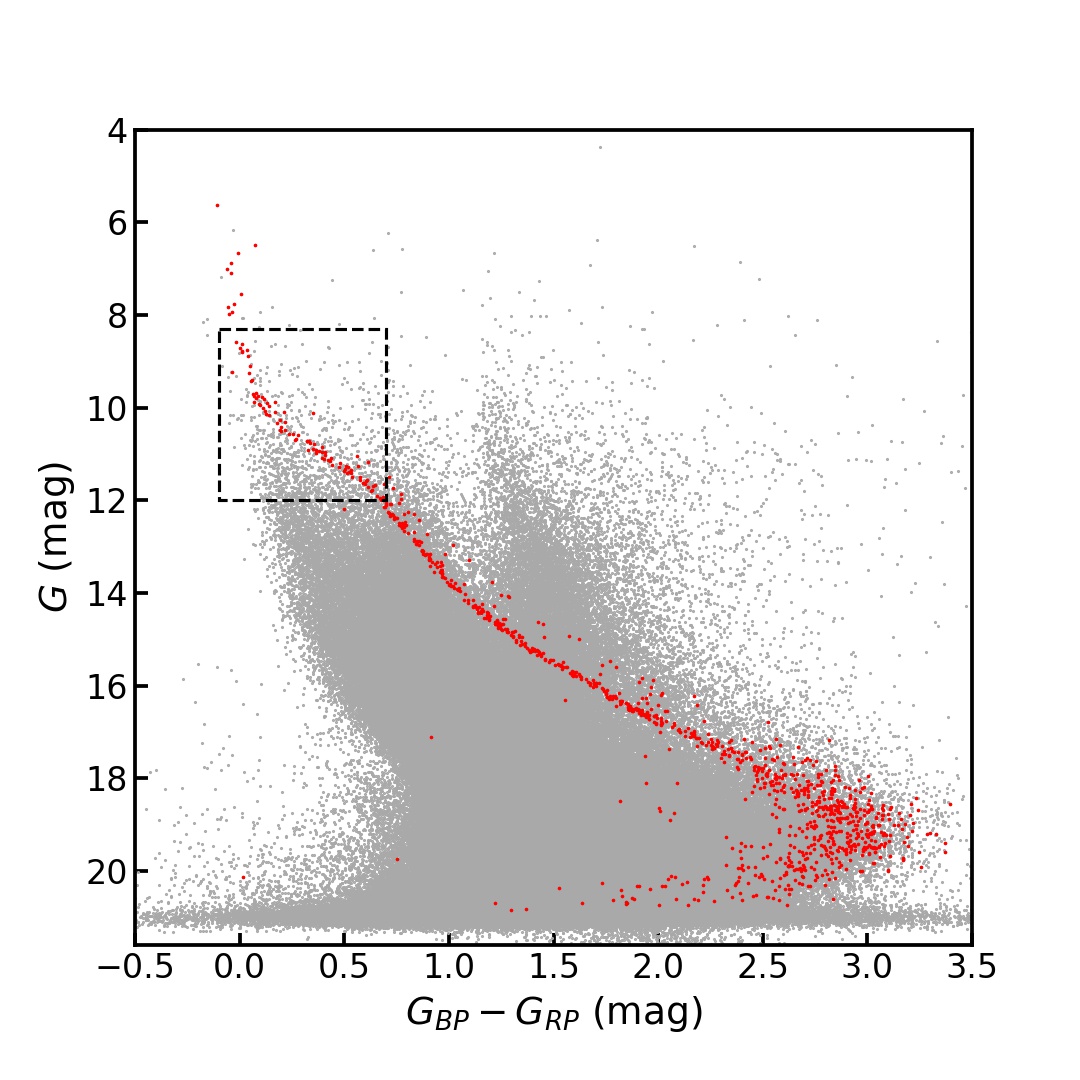}
   \put(80,80){(c)}
  \end{overpic}}
 \hspace{0.001\linewidth}
 \subfigure{\label{fig:cmd_s}
  \begin{overpic}[scale=0.47]{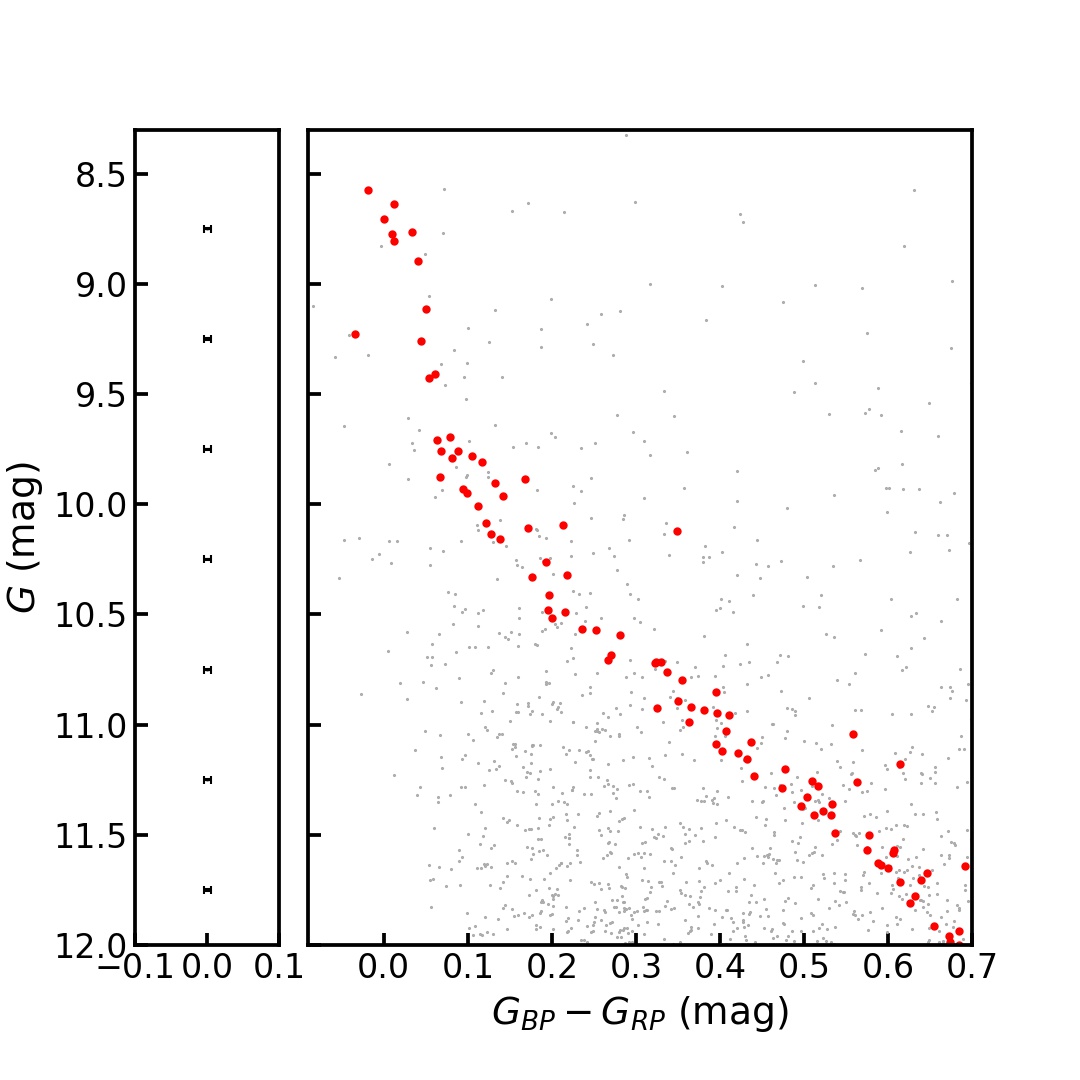}
   \put(80,80){(d)}
  \end{overpic}}

\caption{(a) Selection of NGC 2422 cluster members based on their
  stellar proper motions (see the text for details). The inset shows
  the isodensity contours at $\rho=\unit[1000]{yr^2\,mas^{-2}}$
  (black) and $\rho=\unit[2000]{yr^2\,mas^{-2}}$ (red) of stars within
  the green dashed circle. (b) Distribution of the parallaxes
  ($\varpi$) as a function of \textit{Gaia} $G$-band magnitudes of
  cluster members (red dots), candidate members selected only based on
  proper motions (blue dots), and field stars (grey dots). The
  corresponding number distributions are shown using the same colors
  in the inset, except for the field stars. The dashed polygon
  encloses the $\varpi$ range of the cluster members. The solid and
  dashed vertical lines in the inset show $\varpi_{a}$ of the cluster
  members and the range of $\varpi_{a}\pm \sigma_{\varpi}$,
  respectively. (c) CMD of the member stars (red dots) and the field
  stars (grey dots). (d) Magnified view of the CMD within the dashed
  box in panel (c), where the split pattern is most evident. The left
  subpanel shows the mean uncertainties in the $G$ magnitudes and
  $G_{\rm BP}-G_{\rm RP}$ colors in the corresponding $G$ magnitude
  ranges.}
\end{figure}

We obtained \textit{Gaia} EDR3 astrometric and photometric
measurements of stars within $\unit[2]{^{\circ}}$
\citep[$\approx$16.87 pc, for a distance to NGC 2422 of $\sim$483
  pc;][]{2018A&A...616A..10G} of the center of NGC 2422 \citep[center
  coordinates from][]{2020A&A...633A..99C}. The reference epoch for
\textit{Gaia} EDR3 is J2016.0. We determined whether they were NGC
2422 members by examining their stellar proper motions
($\mu_{\alpha}\cos\delta, \mu_{\delta}$) and parallaxes ($\varpi$). In
Fig.~\ref{fig:pm}, the diagram of stellar proper motions shows that
NGC 2422 members are well separated and concentrated at
$(\mu_{\alpha}\cos\delta, \mu_{\delta})\approx
\unit[(-7.0,1.0)]{mas\,yr^{-1}}$ \citep{2018A&A...616A..10G}. We
calculated the number density of stars, $\rho$, in Fig.~\ref{fig:pm}
in grids of $(\unit[0.1]{mas\,yr^{-1}})^2$ for stars within
$\sqrt{(\mu_{\alpha}\cos\delta-(-7.0))^2+(\mu_{\delta}-1.0)^2}<\unit[1.5]{mas\,yr^{-1}}$
(green dashed circle in Fig.~\ref{fig:pm}). The average proper motion
of the stars (green dots in the inset of Fig.~\ref{fig:pm}), enclosed
by the isodensity contour of $\rho=\unit[2000]{yr^2\,mas^{-2}}$ ,
$(\mu_{\alpha}\cos\delta, \mu_{\delta})=
\unit[(-7.06,1.02)]{mas\,yr^{-1}}$, was used as the center for
selecting cluster members based on their proper motions \footnote{In the inset of Fig.~\ref{fig:pm}, three populations have $\rho>\unit[2000]{yr^2\,mas^{-2}}$. We only used the most centrally concentrated population with respect to the green dashed circle to determine the circle's center to constrain the proper motions of the member stars. The other two populations were excluded because they are closer to the concentration of the proper motions of the field stars, which may introduce additional contamination}. Based on spectra of 57 FGK stars, \citet{2018MNRAS.475.1609B} reported the
1$\sigma$ dispersion of the radial velocities (RVs) of NGC 2422 stars,
$\unit[0.750\pm0.065]{km\,s^{-1}}$, whose $3\sigma$ dispersion
corresponds to $\unit[0.98^{+0.08}_{-0.09}]{mas\,yr^{-1}}$ for
$\mu_{\alpha}\cos\delta$ and $\mu_{\delta}$ at the distance of NGC
2422, if the motions of member stars in different directions are
isotropic. We thus constrained the member stars to have
$\sqrt{(\mu_{\alpha}\cos\delta-(-7.06))^2+(\mu_{\delta}-1.02)^2}<\unit[0.98]{mas\,yr^{-1}}$
(within the blue dotted circle in Fig.~\ref{fig:pm}), which resulted
in a population of 3015 candidate member stars.

\begin{figure*}[ht]
\gridline{\fig{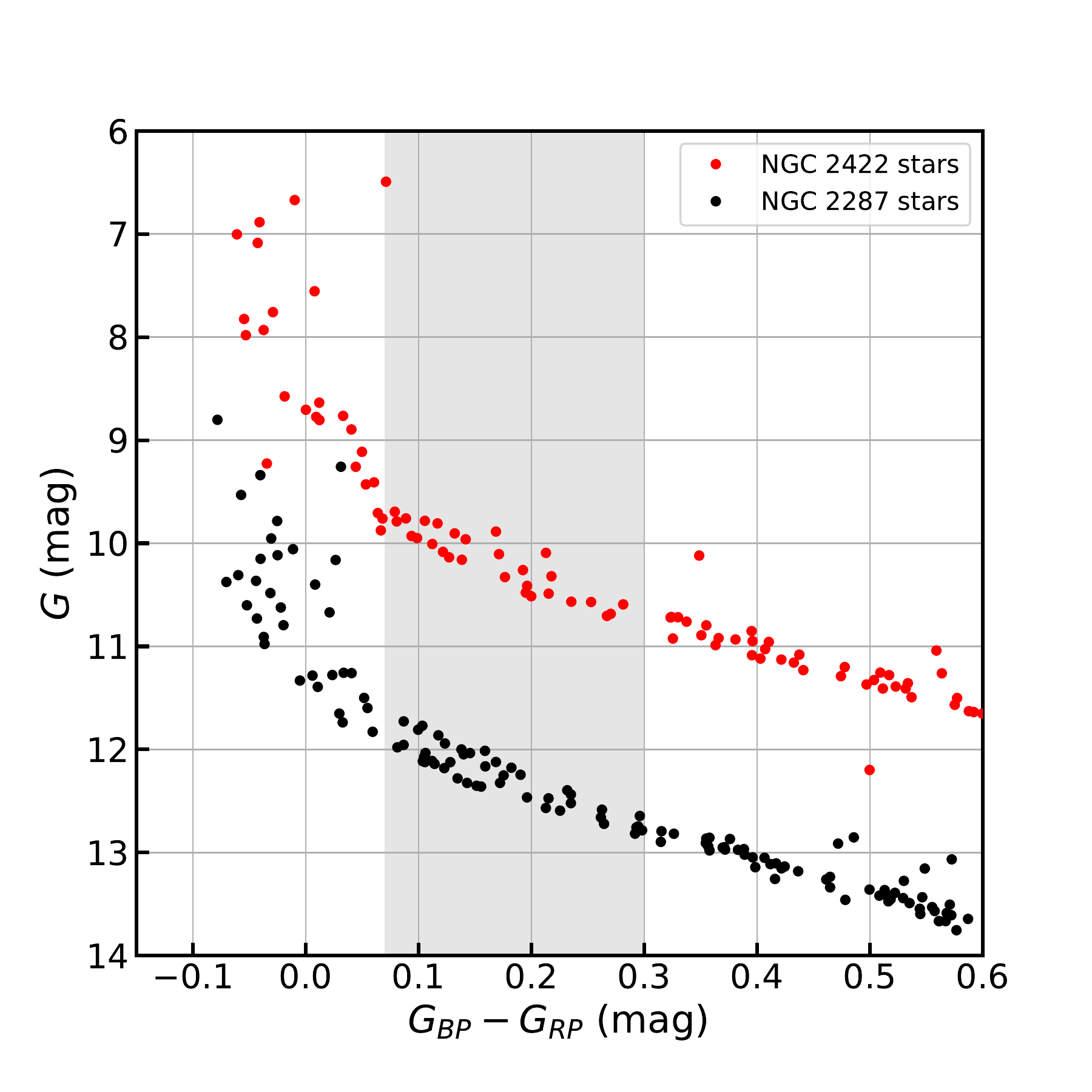}{0.6\textwidth}{}}
\caption{Comparison of the split MS regions of NGC 2422 (red dots) and
  NGC 2287 (black dots). To separate their split MSs, the CMD of NGC
  2287 was shifted up by one magnitude along the ordinate. The grey
  area highlights the region where their split patterns are most
  evident.\label{fig:ngc2422_2287}}
\end{figure*}

We further constrained the cluster members by their stellar
$\varpi$. In Fig.~\ref{fig:para} and its inset, the NGC 2422 members
are concentrated at $\sim\unit[2.1]{mas}$ ($\sim\unit[476]{pc}$),
which is consistent with the $\varpi$ of the cluster from
\citet{2018A&A...616A..10G} ($\sim \unit[2.069]{mas}$,
$\sim\unit[483]{pc}$). We first selected stars that have
$\unit[1.8]{mas}<\varpi<\unit[2.4]{mas}$ (corresponding to a distance
range from \unit[417]{pc} to \unit[555]{pc}) from the 3015 candidate
member stars and calculated the mean ($\varpi_{a}$) and standard
dispersion ($\sigma_{\varpi}$) of their $\varpi$. For stars with
$G<\unit[17]{mag}$, we confined the member stars to have $\varpi$
within $\varpi_{a}\pm \sigma_{\varpi}$, while for fainter stars
($G>\unit[17]{mag}$), $\varpi$ had to be within $\varpi_{a}\pm
3\sigma_{\varpi}$ because of the larger typical parallax measurement
error for fainter stars \citep[][]{2021A&A...649A...1G}. We then
calculated $\varpi_{a}$ and $\sigma_{\varpi}$ for the remaining stars
and repeated this process until the variations in $\varpi_{a}$ and
$\sigma_{\varpi}$ between subsequent iterations were smaller than 1\%
of the mean error in $\varpi$ of stars with $G<\unit[17]{mag}$. We
finally identified 1126 stars as NGC 2422 members. Their average
stellar proper motion and $\varpi$ are $(\mu_{\alpha}\cos\delta,
\mu_{\delta})\approx \unit[(-7.05,1.02)]{mas\,yr^{-1}}$ and
$\unit[2.10]{mas}$ ($\sim\unit[476]{pc}$), respectively, which is
close to that reported by \citet{2018A&A...616A..10G}. The standard
dispersions of the $\mu_{\alpha}\cos\delta$, $\mu_{\delta}$ and
$\varpi$ of the members stars are $\unit[0.33]{mas\,yr^{-1}}$,
$\unit[0.32]{mas\,yr^{-1}}$, and $\unit[0.12]{mas}$,
respectively. Fig.~\ref{fig:cmd_allstar} presents the CMD of selected
cluster members, along with that of field stars. In
Fig.~\ref{fig:cmd_allstar}, the eMSTO and the split MS are visible in
the region for $\textit{G} <\unit[12]{mag}$. Fig.~\ref{fig:cmd_s}
magnifies the loci of the MS within the dashed box in
Fig.~\ref{fig:cmd_allstar} where the split feature is most evident. This pattern is obvious in the $\unit[150]{Myr}$-old LMC cluster NGC 1844 and a Galactic open cluster NGC 2287 as well \citep{2013A&A...555A.143M,2019ApJ...883..182S}. We confirmed that photometric uncertainties were not responsible for the
split pattern of the MS (see Fig.~\ref{fig:cmd_s}). In
Fig.~\ref{fig:ngc2422_2287}, we show the split MS regions in NGC 2422
and NGC 2287. The widths of the split features in both clusters are
comparable.

\subsection{Isochrone Fitting\label{sec:iso fitting}}

\begin{figure*}[ht]
 \centering
 \subfigure{\label{fig:iso_fit}
  \begin{overpic}[width=0.8\linewidth]{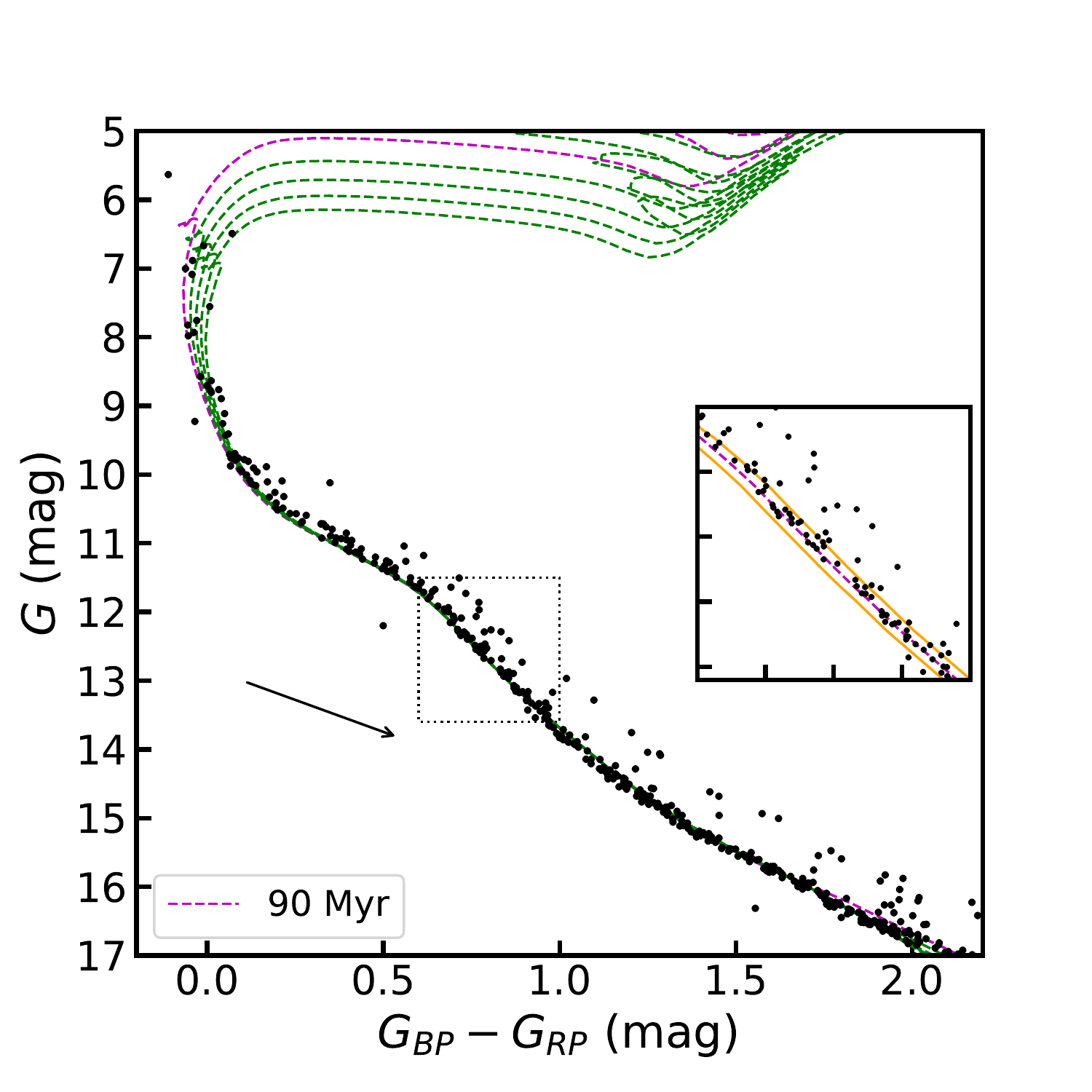}
  \end{overpic}}
          
\caption{CMD of NGC 2422 member stars (black dots) based on
  \textit{Gaia} EDR3. Isochrones of ages ranging from \unit[90]{Myr}
  to \unit[170]{Myr} in steps of \unit[20]{Myr} are shown as dashed
  lines. The magenta dashed line shows the best-fitting isochrone to
  the blue edge of the eMSTO, with an age \unit[90]{Myr},
  $A_\mathrm{V} = \unit[0.3]{mag}$, $Z=0.0133$, and a distance modulus
  $(m-M)_{0}=\unit[8.29]{mag}$. Other isochrones are shown as green
  dashed lines. The black arrow represents the reddening vector,
  corresponding to $\Delta A_\mathrm{V}=1.0$ mag. The inset shows the
  MS region used to test for differential reddening, where the solid
  orange lines are obtained by shifting the best-fitting isochrone
  (dashed magenta line) by $\pm\unit[0.02]{mag}$ in
  color. \label{fig:isofit}}
\end{figure*}

We used isochrones from the PARSEC model \citep[version
  1.2S,][]{2017ApJ...835...77M} to fit the CMD of NGC 2422. In
Fig.~\ref{fig:isofit}, the CMD of NGC 2422 is fitted by a set of
isochrones for metallicity $Z=0.0133$ ($Z_{\odot}=0.0152$ in the
PARSEC model) and extinction $A_\mathrm{V} = \unit[0.3]{mag}$ for a
distance modulus
$(m-M)_{0}=\unit[8.29]{mag}$. \citet{2018MNRAS.475.1609B} obtained
[Fe/H] = $-0.05\pm0.02$ dex for NGC 2422, corresponding to $Z =
0.0135\pm0.0006$, which is similar to the metallicity of the
isochrones in Fig.~\ref{fig:isofit}. Their distance modulus
corresponds to a distance of $\sim\unit[455]{pc}$, which is close to
the distance inferred from the average parallax of the cluster
members. The extinction coefficients for the \textit{Gaia} bands were
derived following \citet{1989ApJ...345..245C}, adopting the
\citet{1994ApJ...422..158O} $R_V = 3.1$ extinction curve. The ages of
the isochrones range from \unit[90]{Myr} to \unit[170]{Myr} in steps
of \unit[20]{Myr}, matching an age spread of $\unit[80]{Myr}$ to fit
the width of the MSTO. The best-fitting isochrone to the blue edge of
the MSTO region has an age of $\unit[90]{Myr}$, identified by visual
inspection. The age of NGC 2422 in the literature spans a wide
range. \citet{2013A&A...558A..53K} derived an age of \unit[132]{Myr},
while \citet{2001A&AT...20..607L} gave an age of only
\unit[73]{Myr}. This difference in age estimates for NGC 2422 is
expected. Because of the presence of an eMSTO, the absence of a
red-giant branch also complicates the determination of an accurate
isochronal age. Differential reddening may affect the morphology of
the MSTO region as well \citep[][]{2012ApJ...751L...8P}. To examine
the effect of differential reddening, we shifted the best-fitting
isochrone by $\pm\unit[0.02]{mag}$ in color (see
Fig.~\ref{fig:isofit}), which correspond to the three times the
measurement errors in the stellar colors for stars within the inset of
Fig.~\ref{fig:isofit}. As shown by that inset figure, most stars on
the MS ridgeline reside within the region limited by the shifted
best-fitting isochrones. This indicates that the effect of
differential reddening on the CMD of NGC 2422 is limited. It is thus
unlikely responsible for the appearance of the eMSTO and the split MS.

\subsection{Spectroscopic Data Reduction}

We obtained spectra of 36 A-type stars in NGC 2422 from
Canada--France--Hawai'i Telescope (CFHT) program 20BS002, observed
with ESPaDOnS with a resolution of $R\approx 68,000$ and a
signal-to-noise ratio (SNR) of $\sim 80$ at $\sim$ \unit[4440]{\AA}. We also observed spectra of 22 A- and F-type stars
with the Southern African Large Telescope
\citep[SALT;][]{2006SPIE.6267E..0ZB} through program 2019-2-SCI-023
with a resolution of $R\approx 4000$ and a SNR of $\sim 200$ at a
wavelength of \unit[4884.4]{\AA}. Eleven stars were observed with
both the CFHT and SALT. In total, we have 47 spectroscopic sample
objects to investigate the $v\sin i$ of stars along the split MS.
Information about their magnitudes and the observation facilities is
summarized in Table~\ref{tab:param}.

\begin{figure*}[ht]
\gridline{\fig{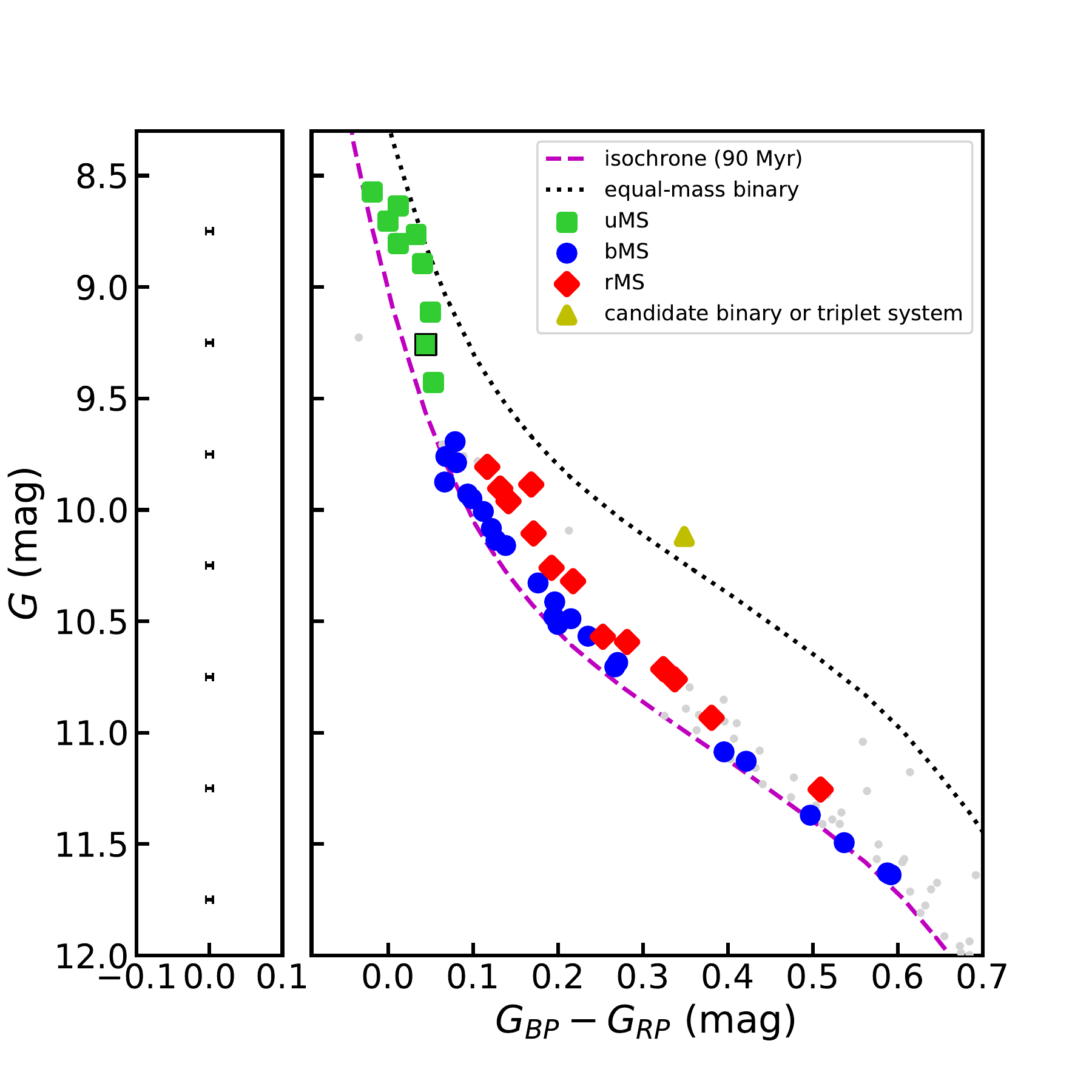}{0.8\textwidth}{}}
\caption{The bMS, rMS, and uMS stars are labeled with solid blue
  circles, red diamonds, and green squares, respectively. The
  candidate binary or triplet system located near the equal-mass
  sequence (the dotted black line) is represented by a solid yellow
  triangle. The double-lined spectroscopic binary system is labeled
  with an open square. The magenta dashed line represents the
  best-fitting isochrone. The light grey dots are other member stars
  for which we do not have spectra. In the left panel, we show the
  mean uncertainties for the $G$ magnitudes and $G_{\rm BP}-G_{\rm
    RP}$ colors of stars in each $G$-magnitude bin of
  $\unit[0.5]{mag}$. \label{fig:cla}}
\end{figure*}

We classified our 47 sample stars into four groups based on their
positions in the CMD (Fig.~\ref{fig:cla}). For stars with
$G>\unit[9.5]{mag}$, the 24 and 13 stars on the left and right sides
of the split MS are classified as bMS and rMS stars, respectively. In
Fig.~\ref{fig:cla}, the equal-mass binary sequence has been plotted by
shifting the best-fitting isochrone by $\unit[-0.75]{mag}$ in the
\textit{Gaia} $G$ band. We identified one star beyond the equal-mass
binary sequence, which may be caused by photometric errors due to
inaccurate point-spread-function fitting residuals, or it may be a
triple system. For stars with $G<\unit[9.5]{mag}$, the bMS, rMS, and
equal-mass ratio binary sequence are too close to be clearly
separated. The 9 stars in this magnitude range are defined as upper MS
(uMS) stars.

\begin{figure}
 \centering
 \subfigure{
  \begin{overpic}[scale=0.3]{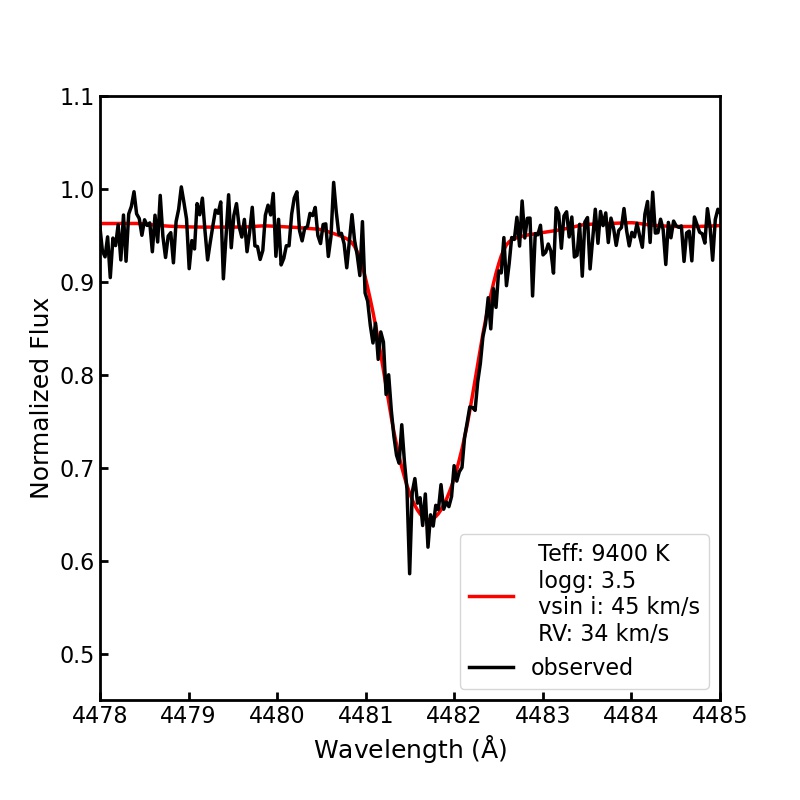}
  \end{overpic}}
  \hspace{0.001\linewidth}
 \subfigure{
  \begin{overpic}[scale=0.3]{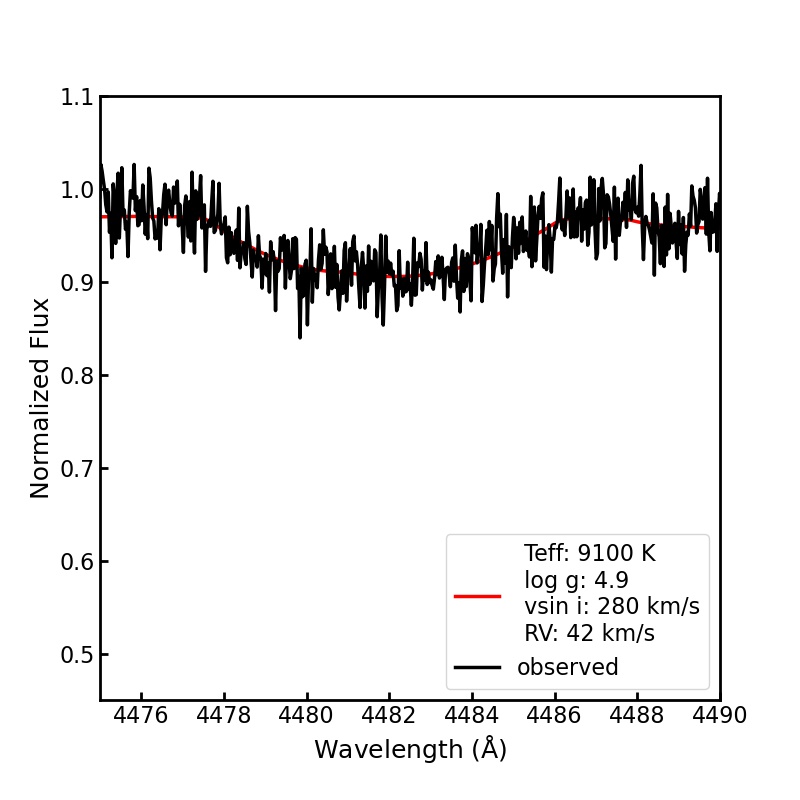}
  \end{overpic}}
  
\caption{Two example spectra of a slowly rotating star (left), and a fast rotating star (right), along with their best-fitting models (red solid lines). Best-fitting parameters are shown in each legend.\label{fig:spectrum samples}}
\end{figure}

To measure the stellar projected rotational velocities, $v\sin i$, we
generated a series of synthetic stellar spectra for a range of
parameters (stellar effective temperatures $T_\mathrm{eff}$, $\log g$,
[Fe/H], $v\sin i$, and RVs) to fit the observed absorption profiles of
the Mg {\sc ii} line ($\sim$ \unit[4481]{\AA}) for the CFHT spectra,
and the $\mathrm{H}_{\beta}$ (\unit[4861]{\AA}) and Mg {\sc i}
triplet (\unit[5100-5200]{\AA}) for the SALT spectra. Synthetic
stellar spectra were derived from the Pollux database
\citep{2010A&A...516A..13P}, with $T_\mathrm{eff}$ ranging from
\unit[6000]{K} to \unit[15,000]{K} (in steps of \unit[100]{K}),
surface gravities from $\log g = 3.5$ to $\log g =5.0$ (in steps of
0.1 dex), and [Fe/H] from $\unit[-1.0]{dex}$ to $\unit[1.0]{dex}$ (in
steps of \unit[0.5]{dex}). Plane-parallel ATLAS12 model atmospheres in local thermodynamic equilibrium were used to generate the
corresponding synthetic spectra \citep{2005MSAIS...8..189K}. A fixed microturblulent velocity of $\unit[2]{km\,s^{-1}}$ was introduced using the tool SYNSPEC \citep{1992A&A...262..501H}. Since the step in [Fe/H] of the derived synthetic spectra is large, we fixed [Fe/H] at $\unit[0.0]{dex}$ in our spectroscopic fitting, based on the metallicity of our best-fitting isochrone (close to $Z_{\odot}$ in the PARSEC model) and the [Fe/H] detected by \citet{2018MNRAS.475.1609B} for NGC 2422 members (see Section~\ref{sec:iso fitting}). We used PyAstronomy \citep{pya} to convolve the synthetic spectra with the effects of instrumental and rotating ($v\sin i$) broadening, and shift the wavelengths based on RVs. The instrumental resolution of CFHT was simulated for the synthetic spectra with a Gaussian kernel of
$\sigma_{c}\approx$ \unit[0.029]{\AA}, whereas that of SALT was
estimated through the width of the corresponding arc line. The $v\sin
i$ values ranged from $\unit[5]{km\,s^{-1}}$ to
$\unit[400]{km\,s^{-1}}$ in steps of $\unit[5]{km\,s^{-1}}$. RVs
ranged from $\unit[0]{km\,s^{-1}}$ to $\unit[70]{km\,s^{-1}}$ in steps
of $\unit[2]{km\,s^{-1}}$, corresponding to the average RV of the 57
NGC 2422 FGK stars ($\unit[35.97\pm 0.09]{km\,s^{-1}}$) measured by
\citet{2018MNRAS.475.1609B}. Then the synthetic flux of the model spectra corresponding to each wavelength value of the observed spectra was calculated with the tool Astrolib PySynphot \citep{2013ascl.soft03023S}. We compared the generated models with the
observed absorption-line profiles and derived the best-fitting
$T_\mathrm{eff}$, $\log g$, [Fe/H], $v\sin i$, and RV for each star
using a minimum-$\chi^2$ method. For the SALT data, we further used a
Monte Carlo method \citep{2013PASP..125..306F} to estimate the
uncertainty in $v\sin i$ around the grid values. In Fig.~\ref{fig:spectrum samples}, we show the spectra of a fast rotating star and a slowly rotating star, along with their best-fitting models.

\begin{figure*}[ht]
\gridline{\fig{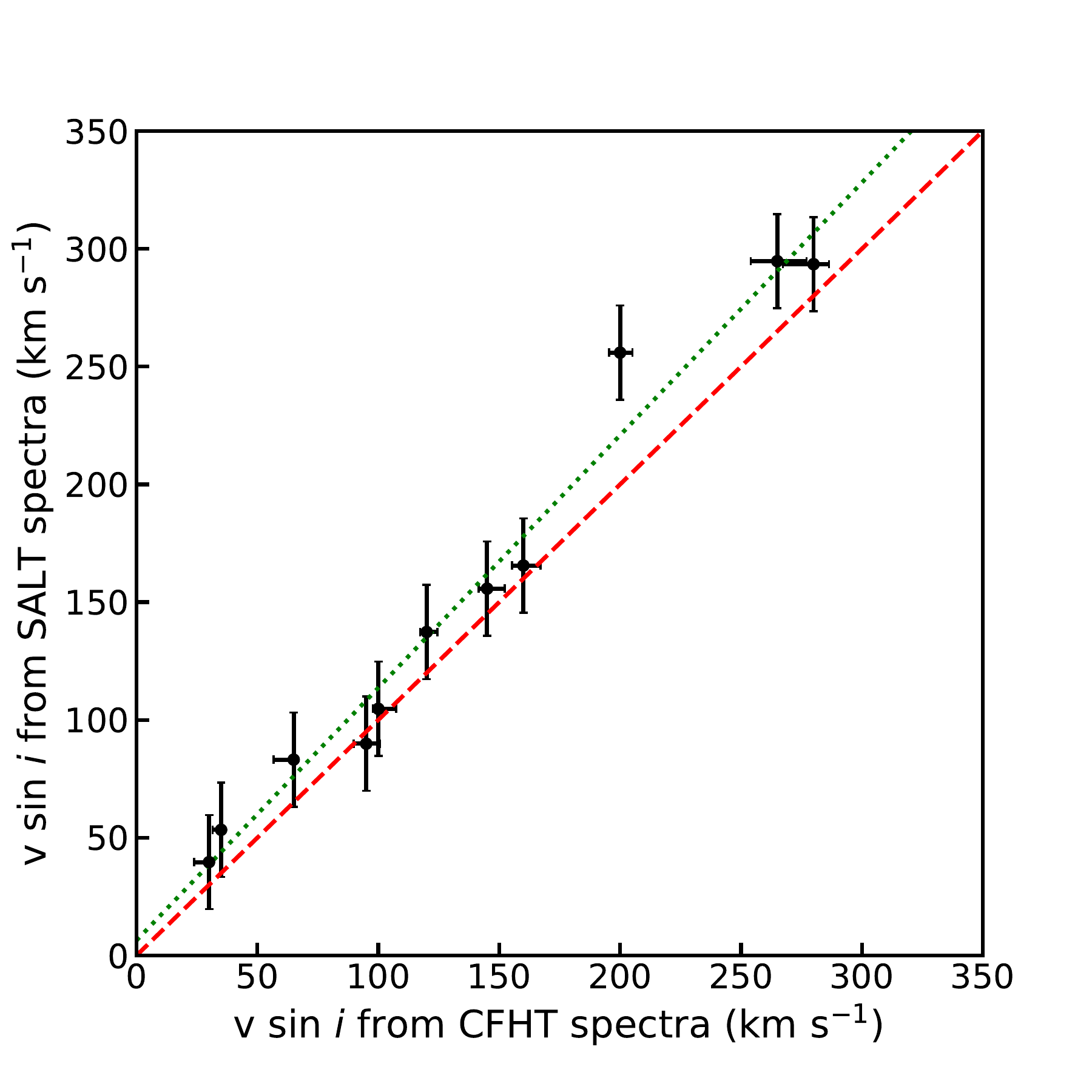}{0.8\textwidth}{}}
\caption{Comparison of derived $v\sin i$ values from CFHT and SALT
  spectra for 11 stars observed with both facilities (black dots). The
  error bars show the $2\sigma$ uncertainties presented in
  Table~\ref{tab:param}. The red dashed line is the equal-value
  track. The dotted green line shows the best-fitting line for the
  data points, representing the function of $v\sin
  i_{\mathrm{SALT}}=1.07 v\sin
  i_{\mathrm{CFHT}}+\unit[6.40]{km\,s^{-1}}$, where $v\sin
  i_{\mathrm{SALT}}$ and $v\sin i_{\mathrm{CFHT}}$ denote the
  projected rotation rates measured from the spectra observed with
  SALT and CFHT, respectively.\label{fig:cfht_salt}}
\end{figure*}

We confirmed that the uncertainties in $T_\mathrm{eff}$ and $\log g$,
which were estimated based on the stars' color--magnitude loci, have
minor effects on their $v\sin i$ determination. Our analysis confirmed
that stars observed by both the CFHT and SALT yielded similar results,
as shown in Fig.~\ref{fig:cfht_salt}. The average of the measured
$v\sin i$ from the CFHT and SALT data was taken as the $v\sin i$
measurement result for each common target. For the data observed with
CFHT, the $v\sin i$ uncertainties were estimated through mock spectra
combined with the $\chi^2$ minimization method
\citep[][]{1976ApJ...210..642A,1996QJRAS..37..519W,2013ApJ...765....4D},
with a 1$\sigma$ uncertainty corresponding to the difference between
the $v\sin i$ values for $\chi^2_{\rm min}$ (the minimum $\chi^2$) and
$\chi^2_{min}$+1. For each common target, we took the uncertainty of
the CFHT data as their $v\sin i$ measurement uncertainty. For stars
observed only with SALT, their average 1$\sigma$ uncertainty of the
$v\sin i$ measurement was $\sim$10 km s$^{-1}$. We included the
$2\sigma$ uncertainties for each star in Table~\ref{tab:param}.

\newpage
\section{Main results and discussion\label{sec:discu}}

\begin{figure*}[ht]
\gridline{\fig{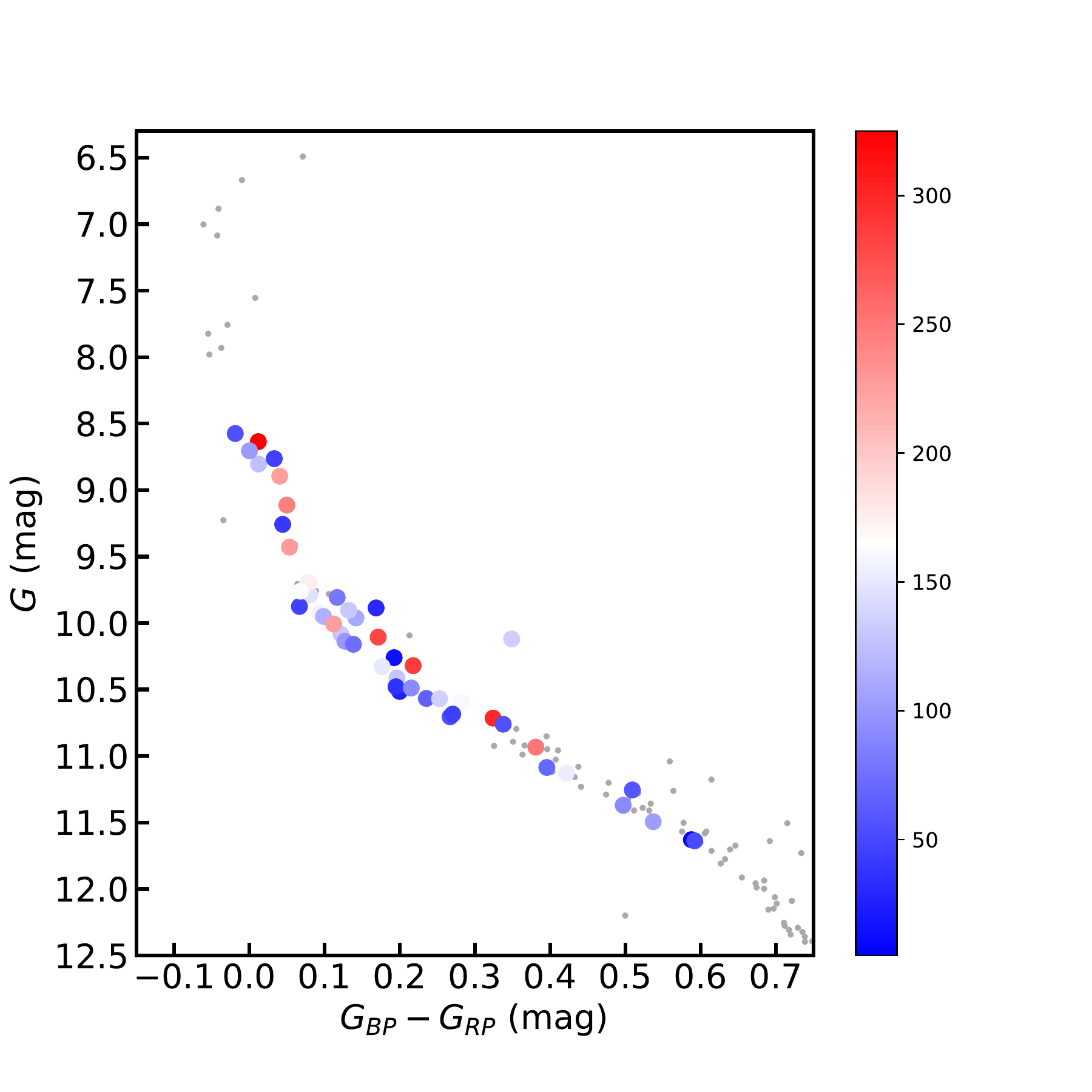}{0.8\textwidth}{}}
\caption{CMD of the 47 stars with $v\sin i$ measurements (color-coded) \label{fig:cmd_vsini}}
\end{figure*}

\begin{figure}
 \centering
 \subfigure{\label{fig:delta_color}
  \begin{overpic}[scale=0.6]{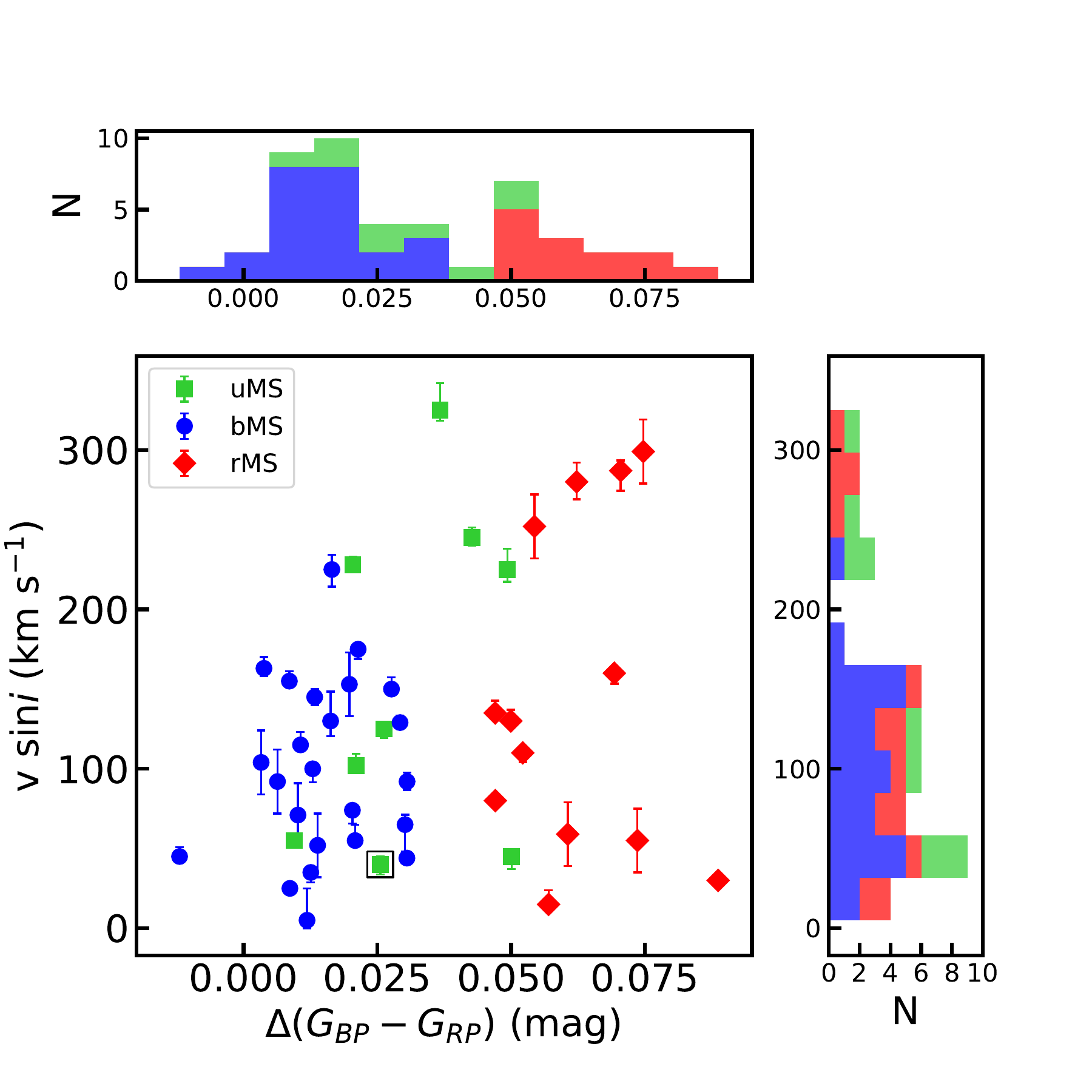}
    \put(80,80){(a)}
  \end{overpic}
}\vspace{0.000\linewidth}
 \subfigure{\label{fig:split_vsini_distribution_cum}
  \begin{overpic}[scale=0.6]{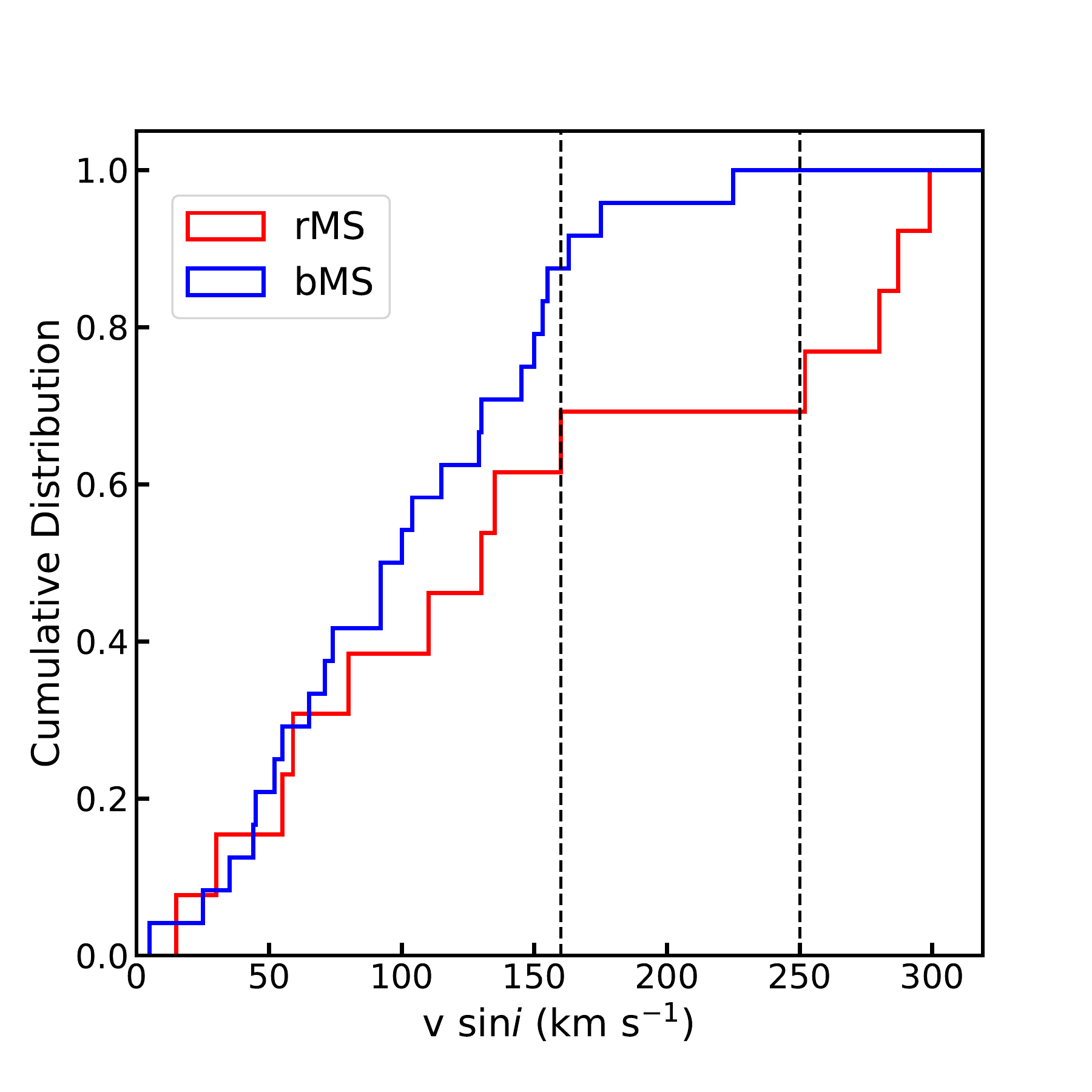}
   \put(80,20){(b)}
  \end{overpic}
}
\caption{(a) Stellar $v\sin i$ (including the $2\sigma$ uncertainties) versus stellar color deviations, $\Delta(G_\mathrm{BP}-G_\mathrm{RP})$, of the bMS, rMS, and uMS stars. The distributions of the color deviations and $v\sin i$ are shown in the top and right panels. The double-lined spectroscopic binary system is enclosed by an open square. (b) Cumulative distributions of $v\sin i$ for bMS and rMS stars. The two dashed vertical lines represent $v\sin i = \unit[160]{km\,s^{-1}}$ (left) and $\unit[250]{km\,s^{-1}}$ (right), respectively.\label{fig:delta_color_vsini_Dist}}
\end{figure}

In Table~\ref{tab:param}, we present the derived projected rotation
velocities for our 47 sample stars with their errors. The distribution
of $v\sin i$ spans the range from $\unit[5.0]{km\,s^{-1}}$ to
$\unit[325.0]{km\,s^{-1}}$. We present the CMD of the 47 stars in
Fig.~\ref{fig:cmd_vsini}, with their $v\sin i$ color-coded. In
Fig.~\ref{fig:delta_color}, we plot the distribution of stellar color
deviations, $\Delta (G_\mathrm{BP} - G_\mathrm{RP})$, versus their
projected rotational velocities, $v\sin i$. The color deviation is
defined as the difference of a star's color ($G_\mathrm{BP} -
G_\mathrm{RP}$) from that of the best-fitting isochrone. In
Fig.~\ref{fig:delta_color}, we have excluded the candidate triple
system. From Fig.~\ref{fig:delta_color} we see that the stellar color
deviation indeed exhibits a bimodal distribution, which again
demonstrates the presence of a bifurcated MS. The $v\sin i$
distribution does not exhibit a bimodal distribution as clearly as
that of the color deviation distribution, although it appears to be
demarcated around $v\sin i\sim \unit[200]{km\,s^{-1}}$. The average $v\sin i$ for bMS stars is $\sim \unit[100]{km\,s^{-1}}$ with a standard deviation of $\sim \unit[54]{km\,s^{-1}}$, and most (23/24) bMS stars have $v\sin
i\leq\unit[175]{km\,s^{-1}}$. The $v\sin i$ distribution of the 13 rMS stars is broader, with a mean and standard dispersion of
$\sim\unit[146]{km\,s^{-1}}$ and $\sim\unit[98]{km\,s^{-1}}$,
respectively. Impressively, 9 out of the 13 rMS stars have $v\sin
i<\unit[160]{km\,s^{-1}}$, rotating with $v\sin i$ like those of most bMS stars, except for 4 rMS stars rotating fast with $v\sin
i>\unit[250]{km\,s^{-1}}$ (see
Fig.~\ref{fig:delta_color_vsini_Dist}). Fig.~\ref{fig:delta_color} does not present any evident correlation between the stellar projected rotation rates and the stellar color deviations for the split-MS stars. The
Spearman correlation coefficient between the color deviation and
$v\sin i$ of the bMS and rMS stars is 0.18, for $p=0.28$, while it is 0.55 \citep{2018ApJ...863L..33M} and 0.68 \citep{2019ApJ...883..182S} for NGC 6705 and NGC 2287, respectively. We further tested the hypothesis that the $v\sin i$ of the rMS and bMS stars of NGC 2422 are generated from the same distribution with the Anderson-Darling test for k-samples. The result shows a significance level of $\sim 0.21$, indicating that the hypothesis can not be rejected. To compare with NGC 2422, we also conducted Anderson-Darling tests for the $v\sin i$ of the bMS and rMS stars of NGC 1818 \citep{2018AJ....156..116M}, NGC 6705 \citep{2018ApJ...863L..33M} and NGC 2287 \citep{2019ApJ...883..182S}. Their tests all reported significance levels smaller than 0.001, implying that the $v\sin i$ distributions of the bMS and rMS stars of each of these three clusters are different. Therefore, the bMS and rMS stars of NGC 2422 are not strongly correlated with slowly and fast rotating populations, respectively, which is very different to the cases in NGC 1818 \citep{2018AJ....156..116M}, NGC 6705 \citep{2018ApJ...863L..33M} and NGC 2287 \citep{2019ApJ...883..182S}.

\startlongtable
\begin{deluxetable*}{CCCCCCcc}
\tablecaption{Photometric information and projected rotational velocities of NGC 2422 stars\label{tab:param}}
\tablewidth{0pt}
\tablehead{
\colhead{\textit{Gaia} ID}&\colhead{$G$ (mag)}&\colhead{$G_\mathrm{RP}$ (mag)}&\colhead{$G_\mathrm{BP}$ (mag)} & \colhead{$\Delta (G_\mathrm{BP} - G_\mathrm{RP})$} & \colhead{$v\sin i$ ($\unit{km\,s^{-1}}$)\tablenotemark{a}} & \colhead{Classification} &\colhead{Facility}\\
\colhead{(1)} & \colhead{(2)} &
\colhead{(3)} & \colhead{(4)} & \colhead{(5)} & \colhead{(6)} & \colhead{(7)} & \colhead{(8)}}
\startdata
3030259479295155072 & 8.57 & 8.55 & 8.57 & 0.009 & $55^{+3.31}_{-2.50}$ & uMS & CFHT\\
3030228413786622720 & 8.64 & 8.63 & 8.61 & 0.037 & $325^{+16.89}_{-6.68}$ & uMS & CFHT\\
3030027447983067008 & 8.70 & 8.69 & 8.69 & 0.021 & $102^{+7.39}_{-2.25}$ & uMS & CFHT \& SALT\\
3030013257412699264 & 8.76 & 8.76 & 8.73 & 0.050 & $45^{+0.21}_{-7.90}$ & uMS & CFHT\\
3029930553522238208 & 8.80 & 8.80 & 8.78 & 0.026 & $125^{+2.61}_{-5.43}$ & uMS & CFHT\\
3030024286886910080 & 8.90 & 8.90 & 8.86 & 0.049 & $225^{+13.01}_{-7.63}$ & uMS & CFHT\\
3029917801751039104 & 9.11 & 9.12 & 9.07 & 0.043 & $245^{+6.30}_{-4.99}$ & uMS & CFHT\\
3028387801268979584 & 9.26 & 9.26 & 9.22 & 0.026 & $40^{+4.98}_{-6.31}$ & uMS & CFHT\\
3030028925451798400 & 9.43 & 9.44 & 9.38 & 0.020 & $228^{+5.13}_{-4.60}$ & uMS & CFHT \& SALT\\
3030681004563366016 & 9.69 & 9.71 & 9.64 & 0.021 & $175^{+4.15}_{-6.09}$ & bMS & CFHT\\
3030035350722750592 & 9.76 & 9.78 & 9.71 & 0.004 & $163^{+7.04}_{-4.74}$ & bMS & CFHT \& SALT \\
3030004530036746368 &9.79 & 9.81 & 9.73 & 0.013 & $145^{+5.18}_{-4.98}$ & bMS & CFHT\\
3030030746517945600 &9.88 & 9.89 & 9.83 & -0.012 & $45^{+5.88}_{-2.41}$ & bMS & CFHT\\
3030228692968781824 &9.93 & 9.96 & 9.86 & 0.009 & $155^{+6.09}_{-3.30}$ & bMS & CFHT\\
3030034732247423360 &9.95 & 9.98 & 9.88 & 0.011 & $115^{+8.13}_{-2.12}$ & bMS & CFHT\\
3030645957629616384 &10.01 & 10.04 & 9.93 & 0.017 & $225^{+9.31}_{-10.68}$ & bMS & CFHT\\
3030219140961660416 &10.08 & 10.12 & 10.00 & 0.016 & $130^{+18.47}_{-9.47}$ & bMS & CFHT\\
3029232707231846784 &10.14 & 10.18 & 10.05 & 0.013 & $100^{+1.07}_{-8.55}$ & bMS & CFHT\\
3030038546178432256 &10.16 & 10.20 & 10.06 & 0.020 & $74^{+1.30}_{-8.32} $ & bMS & CFHT \& SALT \\
3030035522521452032 &10.33 & 10.39 & 10.21 & 0.028 & $150^{+7.33}_{-3.54} $ & bMS & CFHT \& SALT\\
3030027207464910976 &10.41 & 10.48 & 10.28 & 0.029 & $129^{+4.45}_{-2.74}$ & bMS & CFHT \& SALT\\
3030015215917673088 &10.48 & 10.54 & 10.35 & 0.013 & $35^{+0.39}_{-6.14}$ & bMS & CFHT \& SALT\\
3029919592765618304 &10.49 & 10.56 & 10.35 & 0.031 & $92^{+5.61}_{-5.23}$ & bMS & CFHT \& SALT\\
3030069263785446400 &10.51 & 10.58 & 10.38 & 0.009 & $25^{+4.13}_{-0.20}$ & bMS & CFHT\\
3030313802042017536 &10.57 & 10.65 & 10.41 & 0.030 & $65^{+6.22}_{-16.98}$ & bMS & CFHT\\
3030015662586533888 &10.68 & 10.78 & 10.51 & 0.031 & $44^{+0.88}_{-3.52}$ & bMS & CFHT \& SALT\\
3030026588989698048 &10.70 & 10.80 & 10.53 & 0.021 & $55^{+9.90}_{-0.15}$ & bMS & CFHT\\
3030016109262918016 &11.09 & 11.23 & 10.83 & 0.010 & $71\pm20.00$ & bMS & SALT\\
3030014219483828864 &11.13 & 11.28 & 10.86 & 0.020 & $153\pm20.00$ & bMS & SALT\\
3030014013325413248 &11.37 & 11.55 & 11.06 & 0.006 & $92\pm20.00$ & bMS & SALT\\
3030026138007337088 &11.49 & 11.69 & 11.15 & 0.003 & $104\pm20.00$ & bMS & SALT\\
3030025661276778880 &11.63 & 11.85 & 11.26 & 0.012 & $5^{+20}_{-5}$ & bMS & SALT\\
3030028444415416704 &11.64 & 11.86 & 11.27 & 0.014 & $52\pm20.00$ & bMS & SALT\\
3030231785345188608 &9.81 & 9.84 & 9.72 & 0.047 & $80^{+4.22}_{-3.85}$ &rMS & CFHT\\
3030250751920573696 &9.89 & 9.94 & 9.77 & 0.089 & $30^{+3.70}_{-0.23}$ & rMS & CFHT\\
3029983914193883648 &9.90 & 9.94 & 9.81 & 0.050 & $130^{+6.95}_{-3.18}$ & rMS & CFHT\\
3030033495296943488 &9.96 & 10.00 & 9.86 & 0.052 & $110^{+3.30}_{-6.00}$& rMS & CFHT\\
3030015662592785664 &10.11 & 10.16 & 9.99 & 0.062 & $280^{+12.08}_{-10.95}$ & rMS & CFHT \& SALT\\
3030298374519750912 &10.26 & 10.32 & 10.13 & 0.057 & $15^{+8.76}_{-0.52}$ & rMS & CFHT\\
3030034972765597440 &10.32 & 10.39 & 10.17 & 0.070 & $287^{+6.36}_{-12.66}$ & rMS & CFHT \& SALT \\
3033318767322824704 &10.57 & 10.66 & 10.40 & 0.047 & $135^{+7.72}_{-1.83}$ & rMS & CFHT\\
3030085756457512832 &10.59 & 10.69 & 10.41 & 0.069 & $160^{+3.40}_{-6.71}$ & rMS & CFHT\\
3029932821264851968 &10.71 & 10.83 & 10.51 & 0.075 & $299\pm20.00$ & rMS & SALT\\
3030028684933623808 &10.76 & 10.88 & 10.54 & 0.074 & $55\pm20.00$ & rMS & SALT\\
3030016143629067904 &10.93 & 11.07 & 10.69 & 0.054 & $252\pm20.00$ & rMS & SALT \\
3030022152277677440 &11.26 & 11.44 & 10.94 & 0.061 & $59\pm20.00$ & rMS & SALT\\
3029931068918279296 &10.12 & 10.24 & 9.89 & 0.237 & $134\pm 20.00$ & binary & SALT\\
\enddata
\tablenotetext{a}{The errors shown correspond to $2\sigma$ uncertainties.}
\tablecomments{(1) \textit{Gaia} ID in EDR3; (2, 3, 4) \textit{Gaia} bands; (5) Color deviations; (6) Projected rotational velocities; (7) Stellar classification; (8) Observation facilities.}
\end{deluxetable*}

Taking NGC 2287 as a comparison, NGC 2422 contains a higher fraction
of slowly rotating stars in its rMS, which reduces the correlation
between their $v\sin i$ and color deviations. Only 25\% of the NGC
2287 rMS stars have $v\sin i<\unit[160]{km\,s^{-1}}$
\citep[][]{2019ApJ...883..182S}, while this ratio is over 69\% (9/13)
for NGC 2422. Although most bMS stars are slowly rotating stars, the
presence of the high fraction of stars with low $v\sin i$ is not
consistent with the rotation scenario.

\begin{figure*}[ht]
\gridline{\fig{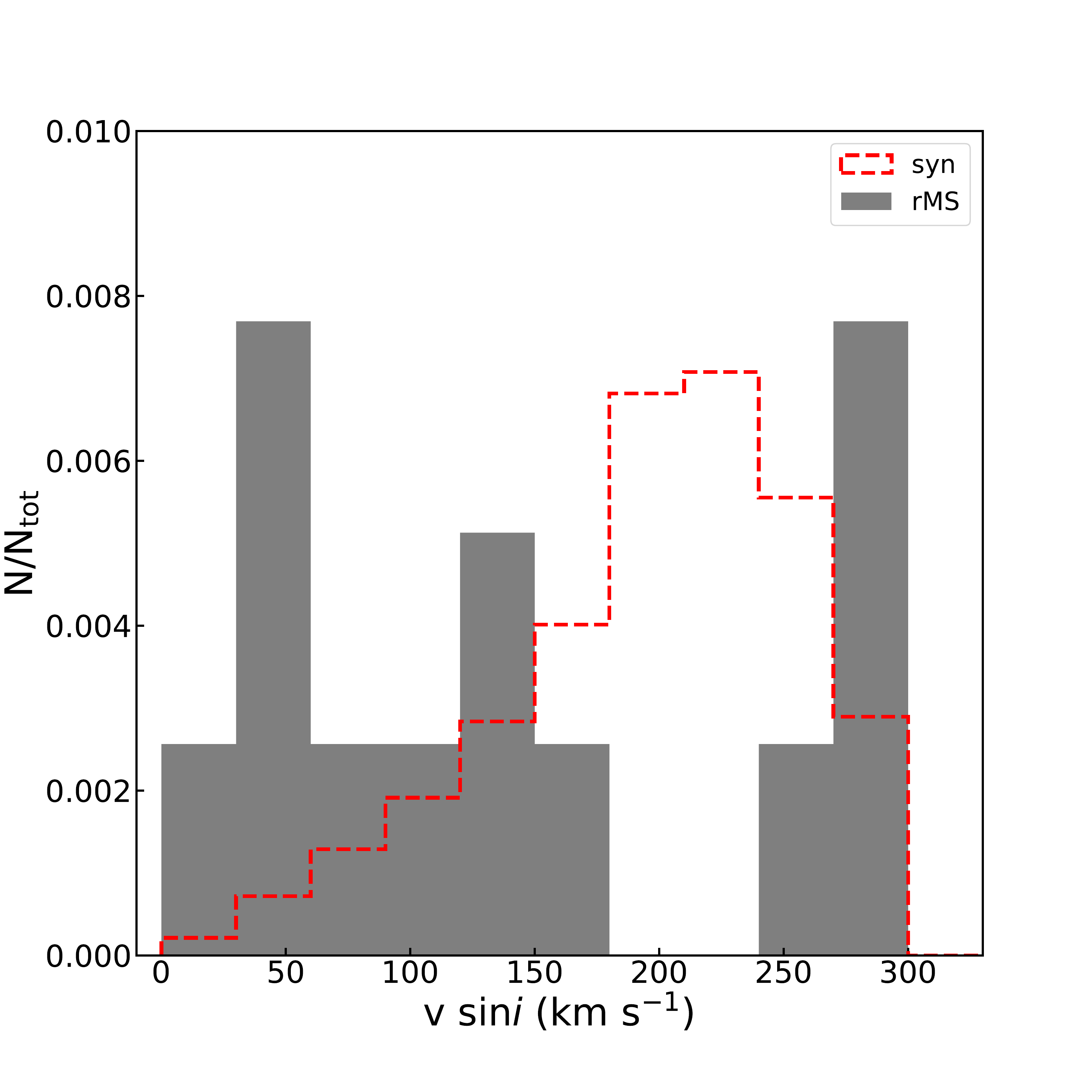}{0.6\textwidth}{}}
          
\caption{Normalized distribution of the $v\sin i$ of 10,000 synthetic
  stars (red dashed line) for a flat distribution of $v$ within the
  range of $\unit[200]{km\,s^{-1}}$ and $\unit[300]{km\,s^{-1}}$, and
  a uniform distribution of the directions of the spin axes in a run,
  along with the $v\sin i$ distribution of rMS stars (grey
  bars)\label{fig:uni_i}}
\end{figure*}

We now discuss the effects of stellar inclinations and binarity. The
projected rotational velocities depend on both the absolute rotational
velocities, $v$, and the stellar inclination, $i$. If we assume that
split-MS stars are single stars, the stellar rotation scenario
suggests that rMS should be populated by fast rotating stars. They may
exhibit small $v\sin i$ values if their inclinations are
small. Slowly/non-rotating stars will never populate the rMS,
however. Based on this consideration, the only explanation which
accounts for the rMS stars with low $v\sin i$ is that their
inclinations are low. Based on the isochrones including stellar
rotation from the SYCLIST database \citep{2013A&A...553A..24G}, we
find that the absolute rotational velocities of stars with $1.7
M_\odot<M<2.5 M_\odot$ (the estimated mass range of the rMS stars) are
about $\unit[210]{km\,s^{-1}}$ to $\unit[280]{km\,s^{-1}}$ at an age
of $\unit[90]{Myr}$. Based on this result, four rMS stars with
$\unit[250]{km\,s^{-1}}<v\sin i<\unit[300]{km\,s^{-1}}$ should rotate
with an edge-on spin-axis. We explored whether a uniform distribution
of spin-axis directions in three-dimensional space can reproduce the
observed $v\sin i$ of the rMS stars. We generated 10,000 synthetic
stars with a uniform distribution of $v$ within
$\unit[200]{km\,s^{-1}}$ and $\unit[300]{km\,s^{-1}}$ and stochastic
orientations of their spin axes in three-dimensional space
\citep[generated from a uniform distribution in $\cos i$
  following][]{2019NatAs...3...76L}. Fig.~\ref{fig:uni_i} shows the
normalized $v\sin i$ distribution of the rMS stars and that of our
10,000 synthetic stars in an example run. We visually excluded the
possibility that stochastic orientations of the spin axes can generate
the observed $v\sin i$. However, we emphasize that small-number
statistics may introduce a large uncertainty. To resolve stellar
inclinations from projected rotational velocities, asteroseismological studies are required to derive their absolute rotational velocities.

Another important factor that may affect our analysis is the presence
of unresolved binaries. To observe a tight correlation between the
stellar rotational velocities and their colors requires that most bMS
stars are single stars or low mass-ratio binaries. Otherwise, they
will contaminate the rMS region, since unresolved high mass-ratio
binaries tend to be located on the red side of the
MS. \citet{2017NatAs...1E.186D} suggested that bMS stars are slow
rotators that spin fast at birth, whose rotation rates are then braked
within 25\% of a cluster's age, and tidal locking in close binaries
could be responsible for this braking. If the tidal-locking scenario
is correct, the photometry of some slow rotators is possibly
contaminated by hidden binary companions that shift their loci in the
red (and bright) direction. These hidden companions may explain the
appearance of the slow rotators in the rMS. Unresolved binaries may
also lead to broadened or split profiles of absorption lines in
stellar spectra. In our spectra, we only found one double-lined
spectroscopic binary system that exhibited an additional absorption
profile at $\sim$ \unit[4479]{\AA} near that of the primary star (at
$\sim$ \unit[4482]{\AA}, \textit{Gaia} ID: 3028387801268979584; for
its position in the CMD and information about its $v\sin i$ and color
deviation, see Fig.~\ref{fig:cla} and Fig.~\ref{fig:delta_color},
respectively), which may indicate the presence of a companion. It has
a slow projected rotational rate ($v\sin
i=\unit[40^{+4.98}_{-6.31}]{km\,s^{-1}}$), which is consistent with
the prediction of the tidal-locking scenario. Intriguingly, this star
was found to be a variable star \citep{2015AN....336..590H}. We did
not detect any clear signatures of unresolved binaries in any of our
other spectra.

We explored if some MS stars may hide a binary companion using the
Binary INformation from Open Clusters Using SEDs (BINOCS) package
\citep{2015AAS...22541503T, 2021AJ....161..160T}. BINOCS was
specifically designed for determining accurate component masses for
unresolved binaries by fitting observed magnitudes from synthetic
stellar SEDs \citep{2015AAS...22541503T,
  2021AJ....161..160T}. Unresolved binaries' SEDs would have an excess
of infrared flux compared with single stars if they hide a low-mass
companion star \citep[][]{2021AJ....161..160T}. In BINOCS, observed
magnitudes from at least three optical, three near-infrared, and two
mid-infrared filters are a prerequisite
\citep{2021AJ....161..160T}. In addition to the \textit{Gaia} $G_{\rm
  BP}$, $G$, and $G_{\rm RP}$ magnitudes, which serve as optical
inputs, for each star we obtained their corresponding 2MASS $JHK_{\rm
  s}$ \citep{2006AJ....131.1163S} and WISE [3.6] and [4.5] passbands
\citep{2010AJ....140.1868W} as our near- and mid-infrared inputs,
respectively. Theoretical stellar SEDs of single and binary stars
(with different mass ratios, $q$) were generated based on the
parameters implied by the best-fitting isochrone from the PARSEC
model. Then, the observed magnitudes were directly compared with the
synthetic stellar SEDs. BINOCS has different $q$ thresholds for binary
stars of different primary masses \citep[][]{2021AJ....161..160T}. In
the mass range explored here, only stars with mass ratios
$\textit{q}>0.3$ are identified as binaries.

We analyzed the binarity for a total of 109 stars. They were composed
of 32 stars from the bMS and rMS stars with $v\sin
i<\unit[200]{km\,s^{-1}}$ and 77 stars located below the split MS
within the range from $\unit[11.50]{mag}<G<\unit[13.60]{mag}$ to
$\unit[0.67]{mag}<G_\mathrm{BP}-G_\mathrm{RP}<\unit[0.98]{mag}$. Both
populations have complete photometric data for the eight input
passbands. We excluded five fast-rotating split-MS stars with $v\sin
i>\unit[200]{km\,s^{-1}}$ from our SED fitting analysis, because the
colors of these stars may be strongly affected by the gravity
darkening effect caused by their fast rotation. BINOCS is not accurate
for the determinaton of the binarity of stars located where the
best-fitting isochrone and equal-mass ratio binary sequence are close
\citep{2021AJ....161..160T}. Therefore, we did not explore the
binarity of the uMS stars based on SED fitting. The loci of the SED-fitting samples in the CMD and the fitting result are shown in
Fig.~\ref{fig:SED samples}.

\begin{figure*}[ht]
\gridline{\fig{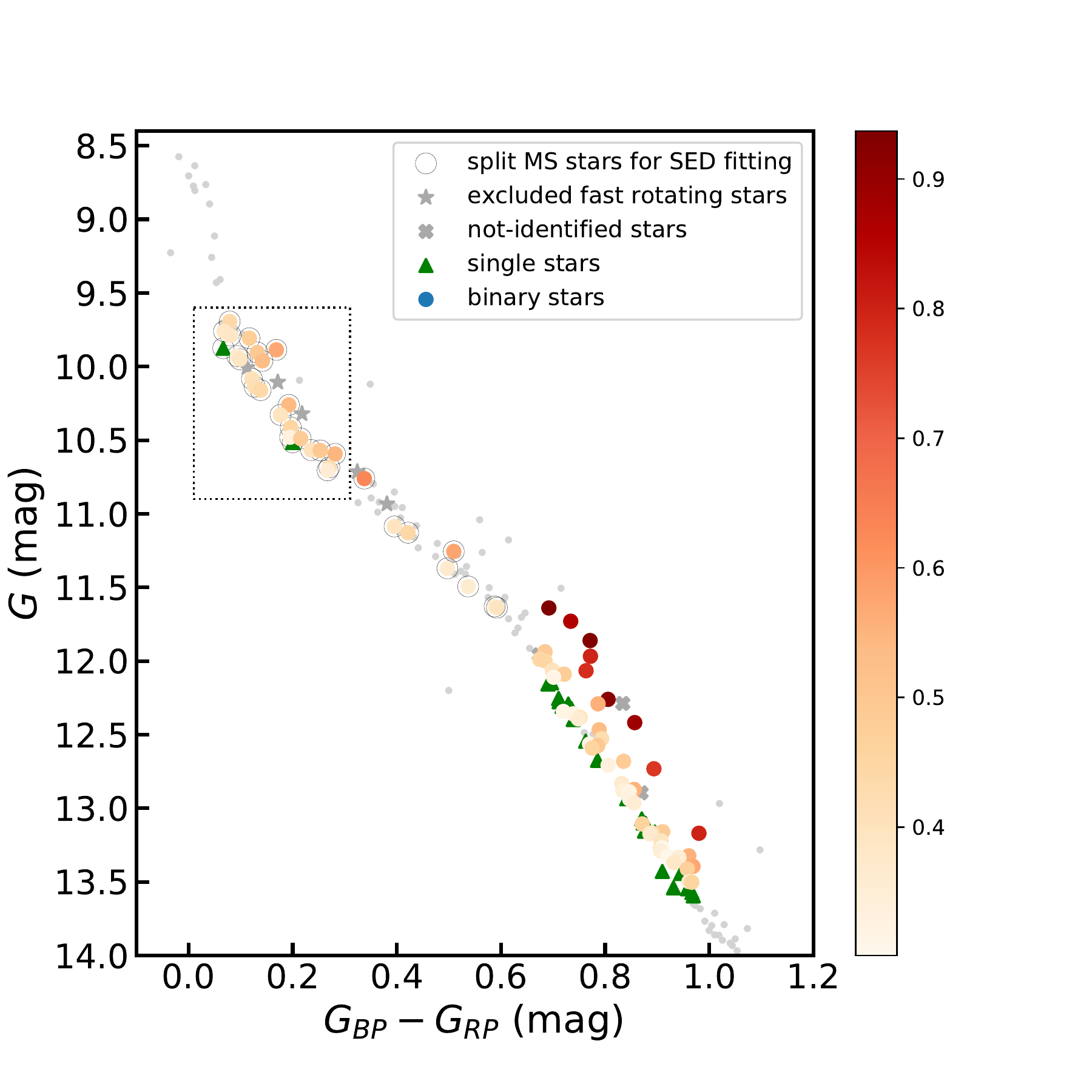}{0.8\textwidth}{}}
          
\caption{CMD of the 109 stars used for our binarity analysis. Stars
  surrounded by open circles are the 32 bMS and rMS stars used for SED
  fitting. Grey star labels show targets discarded from the binarity
  analysis because of their high rotational velocities. The dotted box
  shows the region where 87\% of stars have been analyzed
  spectroscopically. Single and binary stars verified by BINOCS are
  represented by green triangles and solid points, respectively,
  color-coded by the derived mass ratios. Three lower-MS stars with a
  binarity that cannot be determined through SED fitting are marked by
  grey crosses. Small grey dots are other cluster member stars for
  which spectroscopy is not available or which lack complete
  photometric data in the eight prerequisite passbands. \label{fig:SED
    samples}}
\end{figure*}

\begin{figure}
 \centering
 \subfigure{\label{fig:Mass_ratio_distribution}
  \begin{overpic}[scale=0.45]{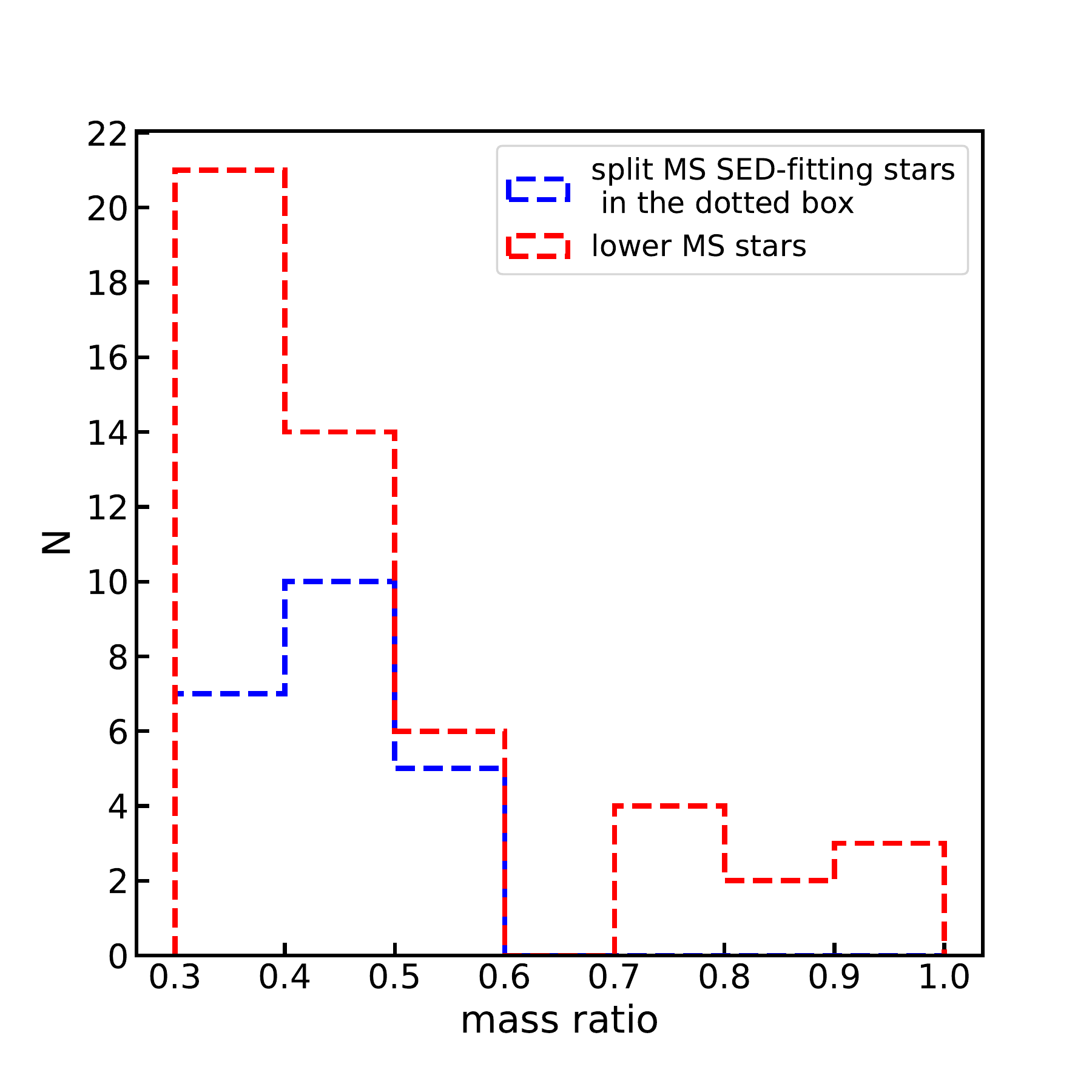}
    \put(80,60){(a)}
  \end{overpic}
}\hspace{0.005\linewidth}
 \subfigure{\label{fig:Split_mass_ratio}
  \begin{overpic}[scale=0.45]{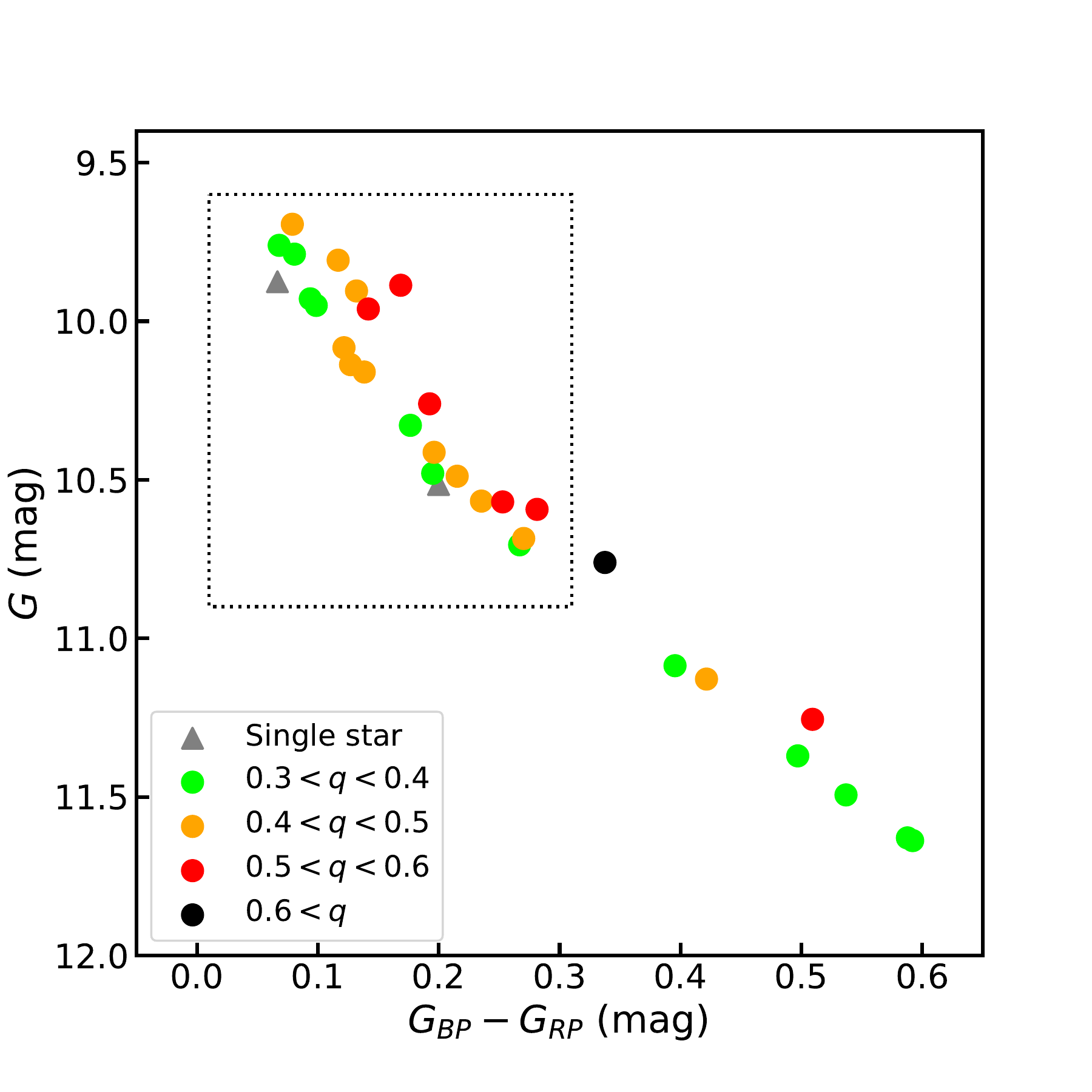}
   \put(80,60){(b)}
  \end{overpic}
}
\caption{(a) Histograms of the binary mass-ratio distributions of the
  sample objects subjected to SED fitting located within the dashed
  box in Fig.~\ref{fig:SED samples} (blue dashed lines) and lower-MS
  stars (red dashed lines). (b) CMD of the 32 split-MS stars used for
  SED fitting, classified based on their derived mass ratios. The
  dotted box shows the same region as that in Fig.~\ref{fig:SED
    samples}.\label{fig:mass_ratio}}
\end{figure}

We determined a binary number fraction for the lower-MS stars of at
least $\sim 65$\% (50/77, while the binarity of three lower-MS stars
cannot be identified; see Fig.~\ref{fig:SED samples}), which is
consistent with the binary fraction of $62\pm16$\% predicted by
\citet{2018MNRAS.475.1609B}. Thirty of the 32 split-MS objects are
identified as unresolved binaries. We further analyzed stars within
the range from $\unit[9.6]{mag}<G<\unit[10.9]{mag}$ to
$\unit[0.01]{mag}<(G_{BP}-G_{RP})<\unit[0.31]{mag}$ (see the dotted
box in Fig.~\ref{fig:SED samples}) where 87\% (27/31) of the members
have been explored spectroscopically. In this range, 22 of the 24
objects were identified as unresolved binaries, indicating a higher
binary number fraction of at least $\sim 81$\% (22/27, if we assume
that the three fast rotators are single stars) along the split MS than
that along the lower MS. Fig.~\ref{fig:Mass_ratio_distribution}
presents the binary mass-ratio distributions for the split and
lower-MS samples. Intriguingly, we find that both the split and
lower-MS stars lack high mass-ratio binaries ($\textit{q}>0.6$). This
may indicate that the mass-ratio distribution of stars in this MS
range favors a low mass-ratio dominated shape, similar to that derived
by \citet{2007A&A...474...77K} for OB associations. We are cautious
about this conclusion, however, because at the distance of NGC 2422,
many wide binary systems in this MS range might have been resolved. In
Fig.~\ref{fig:Split_mass_ratio}, we find that most rMS stars with
$v\sin i<\unit[200]{km\,s^{-1}}$ have mass ratios within the range
$0.4<\textit{q}<0.6$. This indicates that unresolved binaries with
these mass ratios can populate the rMS.

However, we emphasize that this analysis only proves that unresolved
binaries can explain the red colors of some rMS stars. It does not
mean that they are genuine binaries. Strong gravity darkening caused
by fast rotation can alter the distributions of the effective
temperatures on the stellar surfaces and change the stellar
SEDs. Therefore, both varying stellar rotation rates and inclinations
of the spin axes would influence the capability of BINOCS to identify
unresolved binaries from rotating stars. For stars located below the
split MS, this effect is negligible because stars along the lower MS
are usually slow rotators because of the magnetic braking effect
\citep[][]{1967ApJ...150..551K}. To test the effect of rotation on
BINOCS for split-MS stars, we fitted synthetic SEDs of mock stars
($1.7M_\odot<M<2.5M_\odot$) with different absolute rotation rates,
using the theoretical photometry provided by the isochrones for
rotating stars from the SYCLIST database
\citep{2013A&A...553A..24G}. We found that BINOCS could correctly
identify the binarity of stars at a confidence level of at least
$83.1\pm3.6$\% for stars with absolute rotation rates smaller than
$\sim \unit[130]{km\,s^{-1}}$ \footnote{In our fits of the mock SEDs, only three optical passbands (\textit{Gaia} $G_{\rm BP}$, $G$, and $G_{\rm RP}$) and three near-infrared passbands (2MASS $JHK_{\rm s}$) were used because of the limitation of the isochrones. This may decrease the correction ratio for BINOCS to separate single from binary stars.}. We note that about 75\% (24/32) of our sample objects along the split MS have $v\sin i\leq \unit[130]{km\,s^{-1}}$. For the
rMS stars, this ratio is $\sim 78$\% (7/9). This indicates that they
might be slowly rotating stars for which stellar rotation did not
significantly change their SEDs. However, we cannot conclude that they
are genuine binaries at a high confidence level due to a lack of
information about their inclinations. To examine if these stars have
binary origins, time-domain studies of their radial velocities or
light curves are mandatory.

\section{Conclusions\label{sec:conc}}

\cite{2017NatAs...1E.186D} suggested that the population of slowly
rotating stars may hide binary companions. These stars may initially
have been fast rotating stars that have spun down due to early-stage
tidal interactions. Assuming that their hypothesis can account for the
presence of slowly rotating stars in NGC 2422, we estimated the
synchronization timescales, $\tau_\mathrm{sync}$, for 10
slowly-rotating stars with $v\sin i\leq \unit[100]{km\,^{-1}}$ within
the dotted box in Fig.~\ref{fig:SED samples}. Following
\cite{2002MNRAS.329..897H}, we estimated $\tau_\mathrm{sync}$ for
stars with radiative envelopes,
\begin{equation}
	\frac{1}{\tau_\mathrm{sync}}=52^{5/3}\left(\frac{GM}{R^3}\right)^{1/2}\frac{MR^2}{I}q^2(1+q)^{5/6}E_2\left(\frac{R}{a}\right)^{17/2},
\end{equation} 
where $G$ is the gravitational constant and $M$ and $R$ are the
primary star's mass and radius, respectively. $a$ is the separation of
the binary components, $q$ denotes the mass ratio, and $I$ is the
moment of inertia. The units used in the equation are in the CGS
system. $E_2$ is a second-order tidal coefficient that can be fitted
to values given by \citet{1975A&A....41..329Z},
\begin{equation}
    E_2=1.592\times10^{-9}M^{2.84},
\end{equation} 
where $M$ is in units of $M_{\odot}$. To estimate the synchronization
timescales, we used $M$ and $q$ derived from the SED fitting. We
estimated $R$ based on the empirical relation, $R\approx
1.33\times{M}^{0.555}$ for stars with $M>\unit[1.66]{M_\odot}$
\citep{1991Ap&SS.181..313D}.

\begin{figure*}[ht]
\gridline{\fig{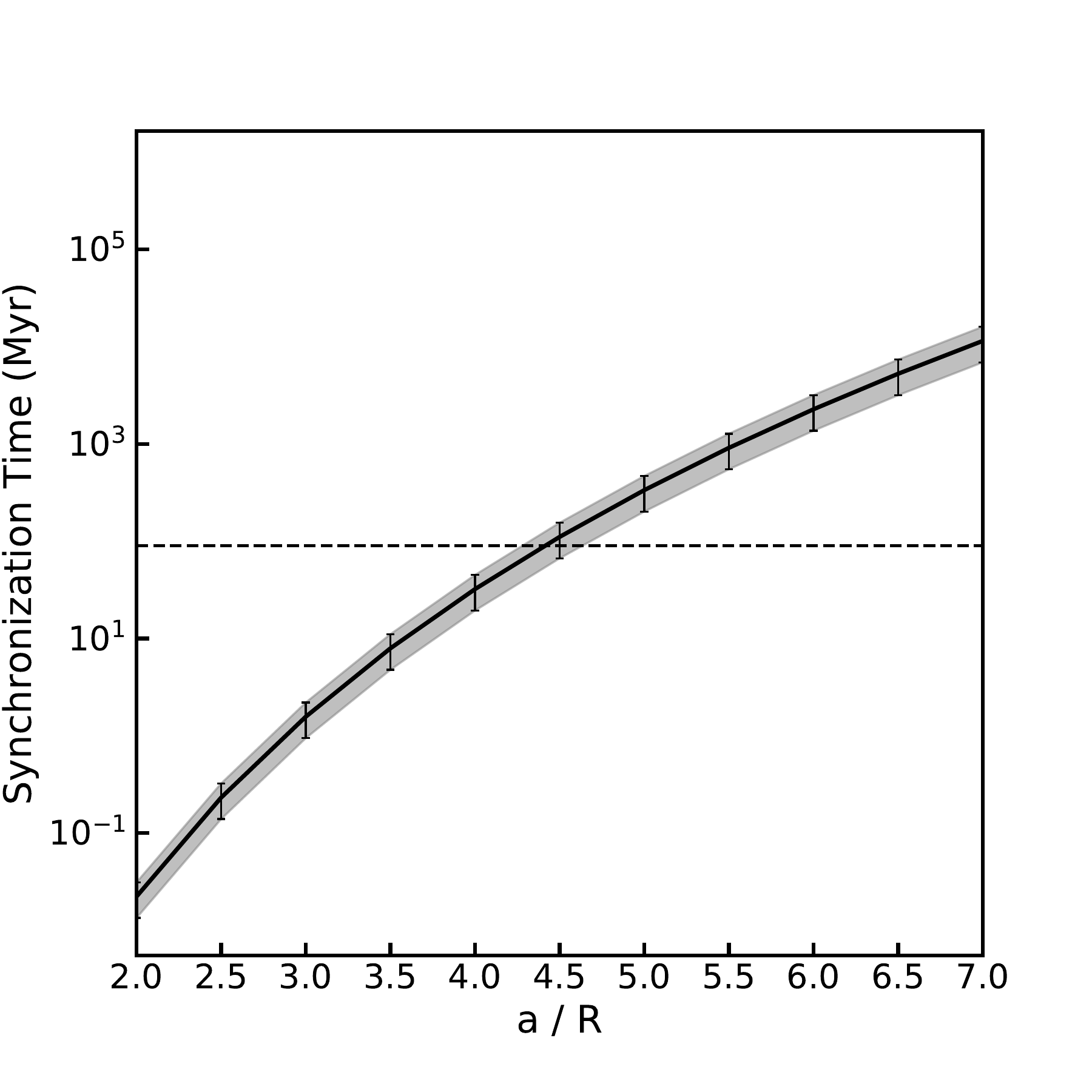}{0.6\textwidth}{}}
          
\caption{Mean synchronization timescales versus different binary
  separations $a/R$ for the 10 slowly rotating stars ($v\sin
  i<\unit[100]{km\,^{-1}}$) located within the dotted box of
  Fig.~\ref{fig:SED samples}. The grey shade represents the 1$\sigma$
  standard deviation of the synchronization time. The dashed line
  indicates \unit[90]{Myr}, i.e., the best-fitting age of NGC
  2422. \label{fig:synchronization time}}
\end{figure*}

In Fig.~\ref{fig:synchronization time}, we present the calculated
synchronization timescales as a function of $a/R$. Our calculation
shows that the mean synchronization timescale dramatically increases
with increasing binary separation relative to their radii, $a/R$. From
Fig.~\ref{fig:synchronization time}, we can estimate that if most
slowly rotating stars are tidally locked on timescales less than the
cluster's age, their separation would not exceed 4.5 times the primary
star's radius, $a/R<4.5$. This indicates that at least 21\% (10/47) of
our spectroscopically explored sample objects are tidally interacting
binaries, which accounts for 37\% (10/27) of the spectroscopically
explored split-MS stars within the dotted box in Fig.~\ref{fig:SED
  samples}. We estimated the orbital periods, $P$, of the companions
for 10 stars using Kepler's third law, assuming $a/R=4.5$. This gives
$P=2.13^{+0.12}_{-0.08}$ days for these stars. However,
\citet{2017ApJS..230...15M} showed that fewer than 10\% of binaries
with A- and late-B primary stars ($2 M_\odot<M<5 M_\odot$) have
companion periods shorter than 10 days, and such a high fraction of
37\% of close binaries with $P<10$ days is only efficient for O-type
MS stars ($M>16M_\odot$). We thus argue that the high fraction of
tidally interacting binaries detected in the split MS is
unusual. Whether or not these slowly rotating stars hide such compact
companions deserves further investigation.

In summary, we have reported on a Galactic open cluster with a
bifurcated MS, NGC 2422, using photometric data from \textit{Gaia}
EDR3. We measured projected rotational velocities, $v\sin i$, for 47
stars along the split MS of NGC 2422, using high- and
medium-resolution spectra observed with CFHT and SALT,
respectively. We found a weak correlation between $v\sin i$ and color,
which is inconsistent with the stellar rotation scenario. This is
caused by the presence of a significant fraction of slowly rotating
stars among the rMS stars. If these rMS stars do not have a specific
inclination distribution, their red colors can be explained by
unresolved binaries. If the presence of slowly rotating stars is not
primordial, but if they were tidally braked by interacting binaries,
\citep[as suggested by ][]{2017NatAs...1E.186D}, their maximum
separations would not exceed five times their primary star's
radius. This would indicate that 37\% of objects in the split MS are
tidally interacting binaries. Such a high fraction, if it is not
primordial, is unusual when compared with observations. Time-domain
measurements of the RVs of these slow rotators are crucial to reveal
whether NGC 2422 is rich in close binaries.

\begin{acknowledgements}
We acknowledge the anonymous referee and the statistics editor of the 
American Astronomical Society Journals for their very useful suggestions. 
This work was supported by the National Natural Science Foundation of China (NSFC) through grant 12073090. This research was also supported
in part by the Australian Research Council Centre of Excellence for
All Sky Astrophysics in 3 Dimensions (ASTRO 3D), through project
CE170100013. This research uses data obtained through the Telescope Access Program (TAP), which has been funded by the TAP member institutes. We acknowledge the science research grants from the China Manned Space Project with NO. CMS-CSST-2021-A08. L.C. and J.Z. acknowledge support from the National Natural Science Foundation of China (NSFC) through grants 12090040 and 12090042. L.L. and Z.S. acknowledge support from the National Natural Science Foundation of China (NSFC) grant U2031139 and the National Key R$\&$D Program of China grant No. 2019YFA0405501. This work has made use of data from the European Space Agency (ESA) mission {\it Gaia}
(\url{https://www.cosmos.esa.int/gaia}), processed by the {\it Gaia}
Data Processing and Analysis Consortium (DPAC,
\url{https://www.cosmos.esa.int/web/gaia/dpac/consortium}). Funding
for the DPAC has been provided by national institutions, in particular
the institutions participating in the {\it Gaia} Multilateral
Agreement. This research has used the POLLUX database
(http://pollux.oreme.org), operated at LUPM (UniversitÃ©
Montpellier--CNRS, France, with the support of the PNPS and INSU. This
publication has made use of data products from the Two Micron All Sky
Survey, which is a joint project of the University of Massachusetts
and the Infrared Processing and Analysis Center/California Institute
of Technology, funded by the National Aeronautics and Space
Administration and the National Science Foundation. This publication
also makes use of data products from the Wide-field Infrared Survey
Explorer, which is a joint project of the University of California,
Los Angeles, and the Jet Propulsion Laboratory/California Institute of
Technology, funded by the National Aeronautics and Space
Administration. 
\end{acknowledgements}

\software{PARSEC \citep[1.2S;][]{2017ApJ...835...77M}, Astropy
 \citep{2013A&A...558A..33A, 2018AJ....156..123A}, Matplotlib \citep{2007CSE.....9...90H}, SciPy \citep{2020SciPy-NMeth}, SYNSPEC \citep{1992A&A...262..501H}, PyAstronomy \citep[][https://github.com/sczesla/PyAstronomy]{pya}, Astrolib PySynphot \citep{2013ascl.soft03023S}, SYCLIST \citep{2013A&A...553A..24G}, BINOCS \citep{2015AAS...22541503T, 2021AJ....161..160T}, TOPCAT \citep{2005ASPC..347...29T}}

\bibliographystyle{aasjournal}
\bibliography{paper.bib}

\end{CJK*}
\end{document}